\pdfoutput=1
\documentclass[fleqn,usenatbib]{mnras}

\usepackage{newtxtext,newtxmath}

\usepackage[T1]{fontenc}

\usepackage[dvipsnames]{xcolor}
\usepackage[normalem]{ulem}
\usepackage{bm}
\usepackage[utf8]{inputenc}
\usepackage{pgfplots}
\usepackage{tikz}
\usepgfplotslibrary{groupplots}
\pgfplotsset{compat=1.14}
\usepackage{capt-of}
\usepackage{graphicx}
\usepackage{aas_macros}

\usepackage{amssymb}

\usepackage{mathrsfs}

\usepackage{amsmath}
\usepackage{hyperref}
\usepackage{lineno}
\usepackage{subfigure}
\usepackage{widetext}

\hypersetup{colorlinks=true, citecolor=blue}

\title[Probing the Dark Solar System]{Probing the Dark Solar System: Detecting Binary Asteroids with a Space-Based Interferometric Asteroid Explorer}
 
\author[A.G. Sullivan et al.]{
Andrew G. Sullivan,$^{1}$\thanks{E-mail: ags2198@columbia.edu}
Do\u{g}a Veske,$^{1}$
Zsuzsa M\'arka,$^{2}$
Imre Bartos,$^{3}$
Szabolcs M\'arka$^1$
\\
$^{1}$Department of Physics, Columbia University in the City of New York, New York, NY 10027, USA\\
$^{2}$Columbia Astrophysics Laboratory, Columbia University in the City of New York, New York, NY 10027, USA\\
$^{3}$Department of Physics, University of Florida, Gainesville, FL 32611-8440, USA
}

\date{Accepted 2022 March 2. Received 2022 February 18; in original form 2021 November 20}

\pubyear{2022}

\hypersetup{draft}
\begin{document}
\label{firstpage}
\pagerange{\pageref{firstpage}--\pageref{lastpage}}
\maketitle

\begin{abstract}
With the inception of gravitational wave astronomy, astrophysical studies using interferometric techniques have begun to probe previously unknown parts of the universe. In this work, we investigate the potential of a new interferometric experiment to study a unique group of gravitationally interacting sources within our solar system: binary asteroids. We present the first study into binary asteroid detection via gravitational signals. We identify the interferometer sensitivity necessary for detecting a population of binary asteroids in the asteroid belt. We find that the space-based gravitational wave detector LISA will have negligible ability to detect these sources as these signals will be well below the LISA noise curve. Consequently, we propose a 4.6 AU and a 1 AU arm-length interferometer specialized for binary asteroid detection, targeting frequencies between $10^{-6}$ and $10^{-4}$ Hz. 
Our results demonstrate that the detection of binary asteroids with space-based gravitational wave interferometers is possible though very difficult, requiring substantially improved interferometric technology over what is presently proposed for space-based missions. If that threshold can be met, an interferometer may be used to map the asteroid belt, allowing for new studies into the evolution of our solar system.

\end{abstract}


\section{Introduction}
Gravitational wave (GW) astronomy promises to advance in the coming decades and will observe more sources and make more precise measurements \citep{2019CQGra..36n3001B, 2021NatRP...3..344B}. In the last few years, GW observatories utilizing interferometric techniques have proven their ability to detect gravitational radiation from inspiraling binary black holes and neutron stars \citep{abbott2016observation, abbott2017gw170817, abbott2019gwtc, 2021arXiv211103606T}. Interferometric GW detection works by measuring the differential displacement of a test mass (TM) acted on by gravitational  rather than electromagnetic forces. This allows for the direct detection of objects difficult to observe by traditional astronomical means. New observations made via GW interferometers will provide key insight into the structure and history of dynamical environments in the universe \citep{schutz1999gravitational}.

Large-scale optical interferometric astrophysical experiments have thus far studied exotic objects beyond our solar system. Not all structures in the solar system, however, have been completely mapped out by traditional astronomical methods. This includes the asteroid belt. Studies of the asteroid belt can probe the history of the solar system, as the belt's structure suggests possible mechanisms for planetary migration and evolution \citep{2010Icar..208..518C, demeo2014solar, 2015MNRAS.453.3619I}. A number of theories concerning the dynamical development of the asteroid belt have been presented \citep{petit2002primordial, izidoro2016asteroid}. Theorized dynamical mechanisms include the mixing of asteroid groups \citep{wilkening19778, clark1995spectral, demeo2014solar, campins2018compositional}
and depletion of asteroid belt mass \citep{petit2002primordial, Clement_2019}. More detailed observation of the asteroid belt will provide evidence of these mechanisms at work.

Studies including the MIPSGAL and Taurus Legacy surveys from the \textit{Spitzer} Space Telescope \citep{2005AJ....129.2869T, 2009Icar..202..104G, 2015A&A...578A..42R} and the NEOWISE survey \citep{2011ApJ...731...53M} have investigated the size and mass distributions of the asteroid belt. These studies rely on the motion and albedo values of asteroids. Consequently, they are limited by the presence and brightness of an asteroid in the fields of view of the detectors, which leads to bias in the observed population \citep{2002aste.book...71J}. 

Binary asteroids have been observed among both Near-Earth and main belt asteroid populations \citep{margot2002binary, pravec2016binary}. In fact, NASA plans to send a mission to the Near-Earth binary asteroid Didymos in 2022 \citep{miao2018towards}. Binary asteroids should have orbits with periods of approximately a day, which correspond to $\sim10^{-5}-10^{-4}$ Hz frequencies.

In the coming decades, GW detection will expand to space-based interferometers. Not limited by seismic noise, these detectors can achieve {high sensitivities at frequencies} under 10 Hz, allowing for the observations of new GW sources \citep{1999ApJ...527..814A, 2004PhRvD..69b2001S, 2016IJMPD..2530001N}. The space-based detector LISA \citep{2017arXiv170200786A} is planned to begin observing in the 2030s. LISA will be sensitive to signals between $10^{-4}$ and $10^{-1}$ Hz. A number of other space-based GW interferometers may achieve high sensitivities in even lower frequency ranges \citep{2016IJMPD..2530001N, ni2013astrod, sato2017status}. Space-based interferometers will be able to make precise measurements of mid to low frequency ($10^{-7}-10^1$ Hz) binary astrophysical sources \citep{2016IJMPD..2530001N}, opening up a new domain of study in astrophysics. 

It has been asserted that LISA may have the ability to detect both the Newtonian gravitational coupling interactions and Shapiro time delay effects from Near-Earth asteroids \citep{2006CQGra..23.4939V, 2007CQGra..24.3005C, 2009CQGra..26h5003T}. We expand on this concept and explore the possibility of an interferometric observational mission dedicated specifically to observing the gravitational effects of asteroid binaries. We consider the binary asteroid detection capabilities of LISA \citep{2017arXiv170200786A}. Additionally, we propose the hypothetical space-based interferometric detectors Asteroid Explorer (AE) and Inner Asteroid Belt Asteroid Explorer (IABAE) whose spacecraft (S/C) would be placed within the asteroid belt. We determine what sensitivities would be needed for AE and IABAE to perform thorough binary asteroid studies.

In this paper, we submit the groundwork for space-based asteroid explorer interferometers. In Section \ref{sec:space}, we introduce background and concepts behind LISA and space-based GW interferometers. In Section \ref{sec:AsteroidBelt}, we provide a review of relevant asteroid belt science. In Section \ref{sec:aster}, we discuss and analyze the signals from binary asteroid sources and their potential detection by a space-based interferometric device. In Section \ref{sec:designandsensitivity}, we present the sensitivity and performance of our proposed large AE detector optimized for asteroid detection. We also outline the scientific improvements needed to make an interferometric asteroid explorer feasible. In Section \ref{sec:InnerAsteroidbelt}, we discuss the performance of our proposed small IABAE detector. We discuss other GW sources to which AE and IABAE would be sensitive in Section \ref{sec:OtherGWs}. We conclude in Section \ref{sec:conc}.

\section{Space-Based Interferometers}
\label{sec:space}
Space-based GW detectors will be sensitive in much lower frequency ranges than ground-based GW detectors. This is simply because space does not have the seismic and Newtonian noise sources which limit the low frequency sensitivity of ground based GW detectors \citep{2016IJMPD..2530001N}. This will allow for the detection of slowly evolving binary sources whose GW radiation is emitted at longer wavelengths. Space-based interferometers such as LISA are designed to detect dimensionless GW strains on the order of $10^{-20}$ \citep{lisa2000lisa, 2019CQGra..36j5011R}. 

\begin{figure}
    \centering
    \includegraphics[width=\columnwidth]{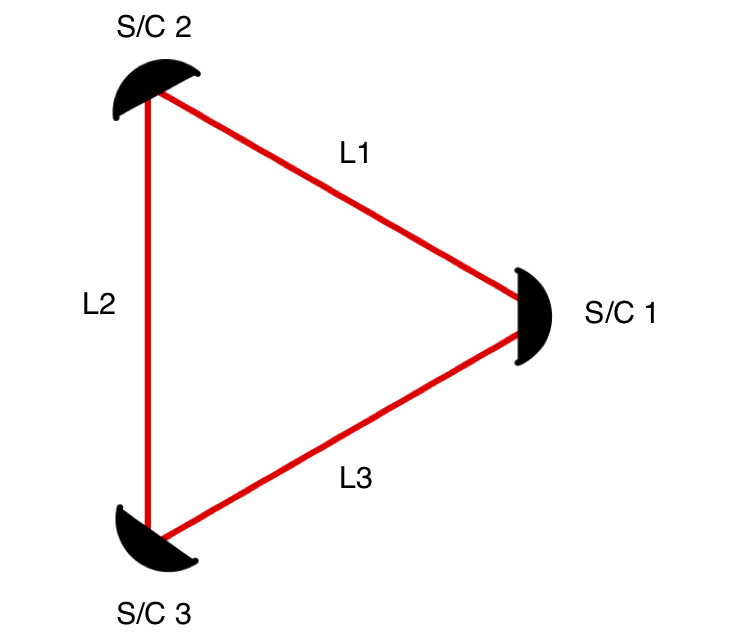}
    \caption{A diagram of the triangular space-based interferometer constellation. The three spacecrafts (S/C) are shown in black while the three laser arms are labeled. The laser arm-lengths vary by detector.}
    \label{fig:constellation}
\end{figure}

LISA and many other space-based detectors will be arranged in nearly equilateral triangles of three S/Cs with on-board TMs. Each side of the triangle will be an interferometric laser arm. Figure \ref{fig:constellation}
shows a basic diagram of a space-based interferometer configuration.  
Unlike ground-based detectors, LISA's TMs will be in free fall motion around the Sun \citep{2017arXiv170200786A}; because of this, the arm-lengths of LISA cannot remain equal and constant as in ground-based detectors. This prohibits exact laser noise cancellation at the photodetectors. As such, LISA and other space-based interferometers will utilize a technique called time delay interferometry to address the delay in laser noise in the interferometer arms \citep{1999ApJ...527..814A, 2004PhRvD..69b2001S, 2014LRR....17....6T, 2021LRR....24....1T}. For data processing, LISA may be treated as an arrangement of three interferometers working in coincidence: two Michelson interferometers sensitive to GWs and one Sagnac interferometer sensitive to non-GW noise sources \citep{2014LRR....17....6T, 2017arXiv170200786A}. 

There are a number of proposed space-based interferometers in the design stage \citep{2016IJMPD..2530001N}. First generation space-based detectors include LISA \citep{lisa2000lisa, 2017arXiv170200786A}, {TianQin \citep{2016CQGra..33c5010L, 2021PTEP.2021eA107M}, Taiji \citep{2020IJMPA..3550075R, 2021PTEP.2021eA108L}}, and potentially ASTROD-GW \citep{2016IJMPD..2530001N, ni2013astrod, 2010cosp...38.3821N, men2010design}. Additionally, proposed detectors with sensitivities in between the frequency regimes of LISA and ground-based detectors include Big Bang Observer (BBO) \citep{harry2006laser} and DECIGO \citep{kawamura2006japanese, kawamura2011japanese}. There is also the recently proposed $\mu$Ares detector which will be sensitive in the $10^{-6}-10^{-4}$ Hz range \citep{2019arXiv190811391S}. {All proposed space-based detectors with the exception of TianQin will have heliocentric orbits as opposed to geocentric orbits \citep{2016CQGra..33c5010L, 2021PTEP.2021eA107M}.}

LISA will have arm-lengths of approximately $2.5\times10^9$ m, while ASTROD-GW will have arm-lengths of $2.6 \times 10^{11}$ m. All three LISA S/Cs will follow the Earth's orbit around the Sun about ten degrees behind, while the S/Cs in ASTROD-GW will be located at 3 Sun-Earth Lagrange points. 

\begin{figure}
    \centering
    \includegraphics[width=\columnwidth]{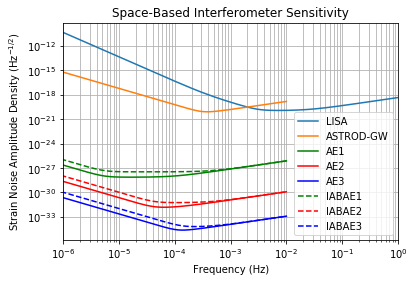}
    \caption{Space-based interferometer sensitivity curves, including LISA, and ASTROD-GW, as well as the AE (See Section \ref{sec:designandsensitivity}) and IABAE detectors (see Section \ref{sec:InnerAsteroidbelt}) proposed in this work. Analytic fits of the strain ASD curves for LISA at L3 sensitivity and ASTROD-GW are plotted using Equation 13 in \protect\cite{2019CQGra..36j5011R} and Equation 14 in \protect\cite{2016IJMPD..2530001N}. The curve for LISA is in blue and the curve for ASTROD-GW is in orange. We use this curve for LISA for the signal-to-noise ratio (SNR) calculations performed in Section \protect\ref{sec:aster}.}
    \label{fig:LISAcurve}
\end{figure}

Figure \ref{fig:LISAcurve} shows the strain noise amplitude spectral density (ASD) curves for LISA \citep{2019CQGra..36j5011R} and ASTROD-GW \citep{2016IJMPD..2530001N} as well as AE and IABAE proposed in this work for comparison.

In 2016, the LISA Pathfinder mission demonstrated the sensitivity capabilities necessary for space-based detectors. LISA Pathfinder achieved the acceleration noise requirements for LISA: the acceleration noise amplitude was shown to be below 10 fm s$^{-2}$/Hz$^{1/2}$ at frequencies greater than 0.1 mHz  \citep{2016PhRvL.116w1101A, PhysRevLett.120.061101}. LISA and the other space-based interferometers will utilize the technology of LISA Pathfinder to achieve their sensitivity goals.  

\section{Review of the Asteroid Belt}
\label{sec:AsteroidBelt}
We now provide a brief review of asteroid belt science including relevant information to our discussion in the sections that follow. For a thorough review of asteroids, see \cite{2015aste.book.....M}, specifically the sections of \cite{2015aste.book...13D} and \cite{2015aste.book..297N}.

The main asteroid belt is situated between Mars and Jupiter. It extends from a semi-major axes of $\sim2.2$ AU to $3.2$ AU in the orbital plane. The asteroid belt is estimated to be $\sim5 \times 10^{-4}$ Earth masses or $3\times10^{21}$ kg, with roughly a third of that mass found in its largest asteroid, Ceres \citep{2002Icar..158...98K}. The number of objects in the asteroid belt with diameters greater than 1 km is estimated to be $1.2 \pm 0.5$ million with millions more smaller objects \citep{2002AJ....123.2070T}. Assuming the asteroid belt to be a torus with inner and outer radii of 2.2 AU and 3.2 AU containing roughly $\sim10^7$ evenly distributed objects, the average particle density of asteroids would be $7.5\times10^{-19}$ km$^{-3}$.  Because of the asteroid belt's low particle density and consequently the low probability of an asteroid collision, a number of missions through the asteroid belt have made successful flybys of asteroids \citep{russell2012galileo, veverka2001landing, glassmeier2007rosetta, brownlee2014stardust, 2011SSRv..163....3R}. The most recent asteroid belt mission was the Dawn Mission which successfully orbited the two largest asteroids in the belt, Ceres and Vesta \citep{2011SSRv..163....3R}. Using Equations 7 and 8 in \cite{1971NASSP.267..595K}, we can compute the probability of an asteroid encounter with a  spacecraft. For a very large spacecraft roughly the size of ASTROD-GW in a circular orbit with a radius of 2.5 AU in the orbital plane and average velocity with respect to asteroids of  4.5 km/s, this particle density yields a probability of an asteroid encounter of 0.0079 over 10 years. Note that 4.5 km/s is approximately the velocity difference between asteroids at the innermost and outermost semi-major axes of the asteroid belt. 

Asteroids are not distributed entirely uniformly due to the gravitational perturbations of the other planets. These perturbations produce what are known as Kirkwood gaps at locations in resonance with Jupiter's orbit \citep{1996CeMDA..65..175M, 2015aste.book...13D}. The Kirkwood gaps divide the asteroid belt into three discernible regions: the inner belt consisting of asteroids whose semi-major axes are less than 2.5 AU, the middle belt consisting of asteroids whose semi-major axes are between 2.5 and 2.8 AU, and the outer belt consisting of asteroids with semi-major axes greater than 2.8 AU. The outer main belt is the most massive portion of the asteroid belt, being 2-10 times more massive than the inner belt \citep{2015aste.book...13D}. 

\begin{figure}
    \centering
    \includegraphics[width=\columnwidth]{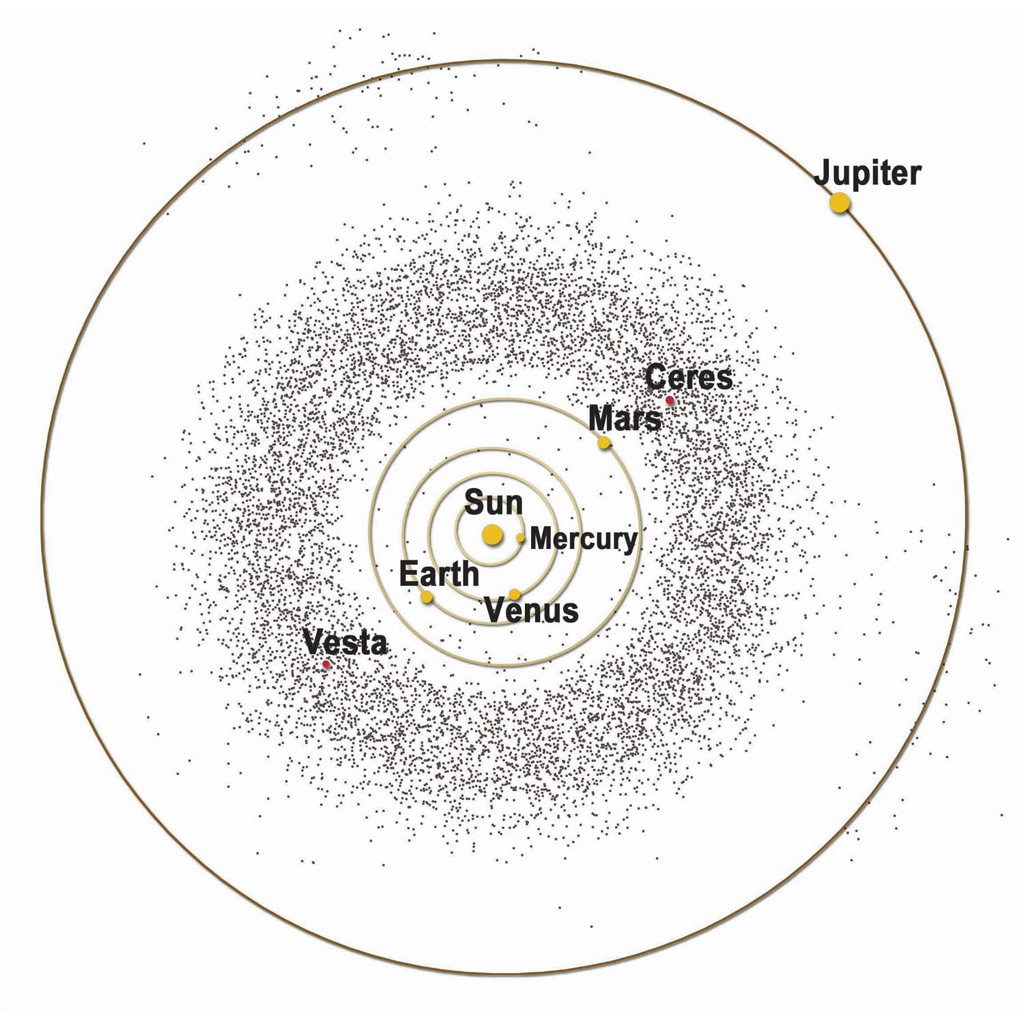}
    \caption{An artist's diagram of the inner Solar System showing the location and orbit of the asteroid belt. The asteroid belt begins outside the orbit of Mars at approximately 2.2 AU and extends to about 3.2 AU. The Sun and planets are labeled as are the two largest asteroids in the belt, Ceres and Vesta. Image taken from NASA's Dawn Mission Art Series \citep{DawnPhotoJ}.}
    \label{fig:AsteroidBElt}
\end{figure}

Orbital eccentricities of asteroids range from 0 to 0.35 while inclinations range from 0 to roughly 30 degrees \citep{2015aste.book...13D}. Figure \ref{fig:AsteroidBElt} shows an overhead diagram of the asteroid belt's orbit.

Asteroids are divided into various spectral classes based on the observable properties of their surfaces. The most populous types of asteroids are S- and C-types \citep{2009Icar..202..160D}. S-type asteroids dominate the inner main belt by mass \citep{2013Icar..226..723D} and are characterized by spectra with moderate silicate absorption at 1 and 2 microns \citep{2015aste.book...13D}.
C-type asteroids, named for connection with carbonaceous chondrite meteorites \citep{2015aste.book...13D}, dominate the outer main belt and make up more than half the main belt by mass \citep{2013Icar..226..723D}. Measurements of asteroids are made by spectral and photometric analyses in the ultraviolet (UV) to infrared range \citep{2015aste.book...13D}.

\begin{figure}
\centering
\includegraphics[width=\columnwidth]{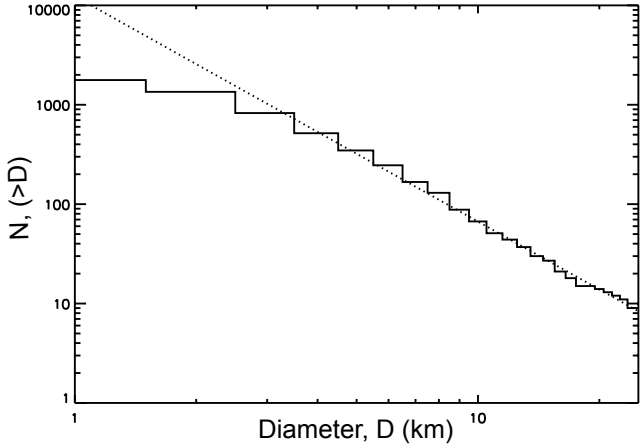}
\caption{Plot of the SFD of the total population of main belt asteroids in the {\textit{Spitzer}} catalogue. The horizontal axis plots asteroid diameter, while the vertical axis plots number. The dashed line represents the power law fit with slope $b={2.34\pm0.04}$. Figure adapted from Figure 11 in {\protect\cite{2015A&A...578A..42R}}.}
\label{fig:AsteroidSFD}
\end{figure}

Asteroid belt science aims at identifying the various observable properties of asteroid populations. Properties include the size frequency distribution (SFD) of the asteroid population \citep{2005AJ....129.2869T,2015A&A...578A..42R} and the average asteroid mass density \citep{britt2003asteroid, carry2012density}. Recent studies of asteroid populations have placed limits on both of these features. Figure \ref{fig:AsteroidSFD} shows the SFD for main belt asteroids obtained by Ryan et al.\ from the \textit{Spitzer} MIPSGAL and TAURUS surveys \citep{2015A&A...578A..42R}. The diameter $(D)$ distribution can be modeled by a power law fit of the form $D^{-b}$ where $b=2.34\pm0.04$ in a limited range. 
The densities of asteroids typically vary by type. S-type asteroids have densities ranging from 2-3 g/cm$^3$ while C-type asteroids have densities ranging from $\sim1.3-2.9$ g/cm$^3$ with the majority having densities below 2 g/cm$^3$ \citep{carry2012density}. However, there remain large uncertainties in both the true SFD and the actual densities of asteroids in the main belt due to difficulty making mass estimates and observational bias.

Properties of the solar system manifest themselves in dynamical interactions between asteroids in the asteroid belt. Collisions are expected to play a dominant role in determining the lifetimes and ages of objects in the asteroid belt and the rest of the solar system \citep{1969JGR....74.2531D, 2005Icar..179...63B}. The average collision rate per cross sectional area of asteroids has been found to be on the order of $\sim10^{-18}$ km$^{-2}$ year$^{-1}$ \citep{1967JGR....72.2429W, farinella1992collision} 
While asteroid collisions have yet to be directly observed, potential collision remnants in the form of debris trails have been found \citep{jewitt2010recent, snodgrass2010collision}. 

Binary asteroids have been observed in the populations of both Near-Earth asteroids and Main-Belt asteroids and represent as many as 16\% of all asteroids \citep{margot2002binary, pravec2016binary, 2015aste.book..375W}. Binary asteroids in the Main Belt have been discovered by both light-curve observation and high resolution imaging techniques \citep{2015aste.book..375W}. Observed binary asteroids are categorized into four major groups by the properties of their primary and secondary asteroids. Large binaries which have primary diameter $D_1>20$ km are in group L and small binaries whose primaries have $D_1<20$ km are in groups A, B, and W \citep{2015aste.book..375W}. The most populous of these small asteroid binary groups is group A \citep{2015aste.book..375W, 2007Icar..190..250P}, characterized by a secondary to primary diameter ratio $D_2/D_1<0.7$ and an orbital separation $d<4.5\times D_1$. Group B consists of binaries with $D_2/D_1>0.7$ and $d<4.5\times D_1$, while group W consists of binaries with $D_2/D_1<0.7$ and $d>4.5\times D_1$ \citep{2015aste.book..375W} The distinguishing characteristic of Group L binaries is their larger primaries \citep{2015aste.book..375W}. Observations of asteroid spin, solar radiation effects, and collision remnants will contain information about the formation of binary systems. In fact, it has also been shown that asteroid collisions may have a non-negligible effect on the lifetimes of binary asteroids \citep{2012MNRAS.425.1492D}. A more thorough catalogue of binaries will allow for precise studies into these dynamical properties.

\section{Asteroid Detection by Space-Based Interferometers}
\label{sec:aster}
\begin{figure}
    \centering
    \includegraphics[width=\columnwidth]{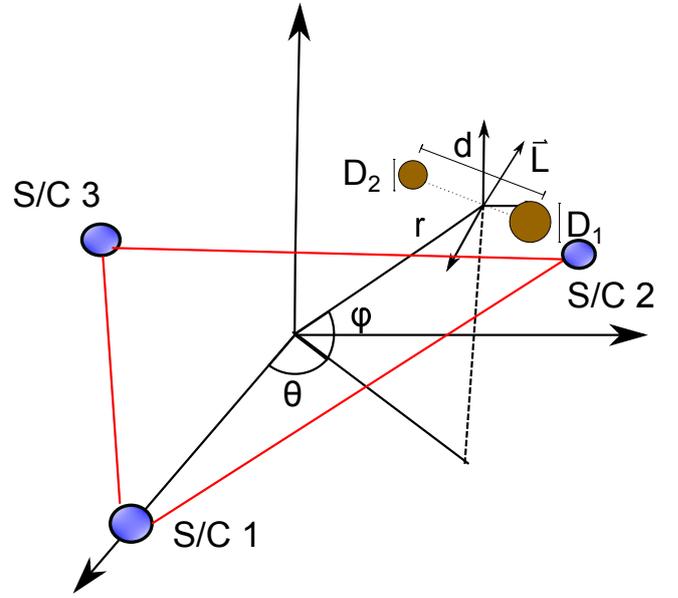}
    \caption{A labeled diagram of the scenario we envision, whereby a three arm space-based laser interferometer is in the vicinity of a binary asteroid system. The two coordinate axes shown are parallel. The interferometer is located in the xy plane and its center is the origin. The coordinates $(r, \theta, \phi)$ denote the position of the center of mass of the binary where $\theta$ is the azimuthal angle and $\phi$ is the altitude angle. $D_1$, $D_2$, and $d$ represent the diameters of the primary and secondary and the binary separation, respectively, while $\vec{L}$ is the orbital angular momentum of the binary.}
    \label{fig:coordinate}
\end{figure}

Because GW interferometers are sensitive to periodic changes in length typically caused by binary astrophysical sources, binary asteroids are a potential target for study. We now analyze the gravitational signals anticipated from binary asteroid systems. Specifically, we study signals associated with the periodic Newtonian gravitational force on the TMs of an interferometer. We assume the binaries to have circular Keplerian orbits and the component asteroids to be spheres. We neglect the spins of the component asteroids. See Figure \ref{fig:coordinate} for the setup. 

Because the masses of a space-based interferometer are in free fall, the acceleration as a function of time of the $i^{\rm th}$ S/C due to the binary can be written in the Newtonian theory as
\begin{equation}
   \vec{a}_i(t)=\sum^{2}_{j=1} \frac{Gm_j}{|\vec{r}_{ij}(r, \theta, \phi, \theta_\ell, \phi_\ell, d, m_1, m_2, t)|^3}\vec{r}_{ij},
   \label{eq: acceleration}
\end{equation}
where $G$ is the gravitational constant, $\vec{r}_{ij}$ is the displacement vector from the $i^{\rm th}$ S/C to the $j^{\rm th}$ asteroid in the binary, $m_1$ and $m_2$ are the masses of the primary and secondary respectively, and $\theta_\ell$ and $\phi_\ell$ are the azimuthal and polar angles, respectively, of the binary orbital angular momentum. $\theta_\ell$ and $\phi_\ell$ are defined with respect to the coordinate axis in Figure \ref{fig:coordinate}. $r$, $\theta$, and $\phi$ denote the location of the center of mass of the binary with respect to the center of the S/C constellation and are assumed to be fixed. $d$ denotes the binary separation. The displacement of each TM is equal to this acceleration integrated twice.

Applying a multipole expansion to Equation \eqref{eq: acceleration} and neglecting the terms without time dependence, we find that the quadrupole term is the periodic term of highest order in distance $r_{i}$ between the $i^{\rm th}$ TM and the center of mass of the binary. Integrating twice, the displacement amplitude of the $i^{\rm th}$ TM in the Newtonian theory will be proportional to 
\begin{equation}
    \text{d}x_i(2\omega)\propto \frac{3G\mu d^2}{ 8\omega^2r_{i}^4},
    \label{eq:2omega}
\end{equation}
where $\mu$ is the reduced mass of the binary, $\omega$ is the orbital angular frequency of the binary. Note that Equation \eqref{eq:2omega} does not include the dependence on angular location and direction of the orbital angular momentum of the binary with respect to the constellation. This term oscillates at a frequency of $2\omega$. 

The TM displacement term oscillating at a frequency of $\omega$ has amplitude proportional to
\begin{equation}
    dx_i(\omega)\propto \frac{3G\mu(m_1-m_2)d^3}{2 M \omega^2 r_{i}^5},
\end{equation}
where M is the total mass. Again, angular location and orbital angular momentum direction dependence are not included.
The $\omega$ frequency term goes as $\propto r^{-5}$ as a result of the symmetry of the binary which cancels the dipole term.   
Consequently, the frequency of interest for signal detection is twice the orbital frequency. The dominant signal in the interferometer data-stream will have an amplitude proportional to the expression in Equation \eqref{eq:2omega} for binaries far from the nearest LISA S/C (i.e.\ $r \gg \frac{(m_1-m_2)d}{ M}$). From this point on we denote twice the orbital frequency as the signal frequency.  

Since the dominant contribution to the signal is the $2\omega$ term, we calculate the displacement amplitude of a TM simply by subtracting the time independent component of the acceleration from Equation \eqref{eq: acceleration} and dividing by a factor of $(2\omega)^2$ to integrate twice. The periodic displacement of the TM from a binary asteroid can thus be approximated as 
\begin{widetext}
\begin{equation}
   d\vec{x}_i(t)\approx\frac{1}{4\omega^2}(\sum^{2}_{j=1} \frac{Gm_j}{|\vec{r}_{ij}(r, \theta, \phi, \theta_\ell, \phi_\ell, d, m_1, m_2, t)|^3}\vec{r}_{ij}-\frac{\omega}{2\pi}\int^\frac{2\pi}{\omega}_0\frac{Gm_j}{|\vec{r}_{ij}(r, \theta, \phi, \theta_\ell, \phi_\ell, d, m_1, m_2, t^{\prime})|^3}\vec{r}_{ij}(t^\prime)dt^\prime).
   \label{eq: displacement}
\end{equation}
\end{widetext}
In order to find the length change of the laser arm, we must take the dot product of $d\vec{x}_i(t)$ from Equation \eqref{eq: displacement} with the unit vectors in the directions of each laser arm. The periodic length changes in the three laser arms become
\begin{subequations}
\label{eq:length}
    \begin{equation}
        \label{eq:lengtha}
       \delta L_1=d\vec{x}_1(t)\cdot \vec{n}_{11}+d\vec{x}_2(t)\cdot \vec{n}_{12},
    \end{equation}
    \begin{equation}
        \label{eq:lengthb}
         \delta L_2=d\vec{x}_2(t)\cdot \vec{n}_{22}+d\vec{x}_3(t)\cdot \vec{n}_{23},
    \end{equation}
    \begin{equation}
        \label{eq:lengthc}
        \delta L_3=d\vec{x}_1(t)\cdot \vec{n}_{31}+d\vec{x}_3(t)\cdot \vec{n}_{33},
    \end{equation}
\end{subequations}
where $\vec{n}_{mi}$ is the unit vector from the $i^{\rm th}$ S/C in the direction of laser arm $L_m$. 

The dimensionless strain in each interferometer due to a binary asteroid signal is
\begin{subequations}
    \begin{equation}
        \label{eq:straina}
        h_1(t)=\frac{\delta L_1-\delta L_2}{L}
    \end{equation}
    \begin{equation}
        \label{eq:strainb}
         h_2(t)=\frac{\delta L_2-\delta L_3}{L},
    \end{equation}
    \begin{equation}
        \label{eq:strainc}
        h_3(t)=\frac{\delta L_1-\delta L_3}{L},
    \end{equation}
\end{subequations}
where $L$ is the length of the laser interferometer arm. We have treated the space-based GW detectors as a composition of three Michelson interferometers for ease of calculation as all three interferometers that make up a space-based detector should be sensitive to non-GW sources.

The GW strain amplitude $h_{GW} \sim \frac{GM(d\omega)^2}{ c^4 r}$ will differ from the Newtonian strain by a factor of $\sim \frac{M\omega^4Lr^3 }{ \mu c^4}$ where M is total mass. Since we have $\omega \lesssim 0.0001$ Hz and $L, r \sim 1$ AU for binary asteroids, we should have $\frac{\omega^4Lr^3} {c^4}\ll 1$, so we do not consider any effects from GWs. In this paper, we study only the Newtonian forces from asteroid binaries. 

\subsection{Near-Earth Binary Detection with LISA}
\label{subsec:Didymos}
As LISA will be in orbit following the Earth, it will be in the vicinity of Near-Earth asteroids. As of the time of this paper's writing, NASA launched the Double Asteroid Redirection Test (DART) mission to the Near-Earth binary asteroid Didymos in November 2021 as part of a planetary defense program to test the capability of probes to redirect potentially dangerous asteroids \citep{2018P&SS..157..104C}. Didymos has an orbital path that comes within 0.04 AUs of that of the Earth, making it an ideal subject of a planetary defense study. DART will make contact with Didymos in Fall 2022 when its distance to Earth will be $11\times10^{6}$ km, and will impart an impulse to Didymos's secondary asteroid to alter its period of rotation by several minutes. Because of this recent interest in Didymos and its close flyby proximity to the Earth, we choose it as a potential candidate for detection with LISA. Didymos is composed of a primary asteroid with a diameter of 780 meters and an orbiting secondary with a diameter of 160 meters. The primary and secondary are separated by approximately 1.18 km. The orbital period of the secondary has been observed to be 11.92 hrs \citep{2020Icar..34813777N}, which by Keplerian analysis implies that the total mass of the system is $5.28\times10^{11}$ kg. 

Due to the non-spherical shape of the primary and secondary of Didymos, one cannot necessarily treat Didymos as a Keplerian sum of point masses; however, higher order effects should contribute to the orbital parameters of Didymos by less than $3\%$ \citep{2020Icar..34913849A}. 
We thus compute the signal by assuming the binary to be composed of two point masses in a circular Keplerian orbit with a period of 11.92 hrs using Equation \eqref{eq: displacement}. To obtain the individual component masses, we assume the mass ratio to be equivalent to the cube of the ratio of their diameters. 
We calculate the amplitude signal to noise ratio (SNR) of each of the dimensionless strain signals calculated with Equations \ref{eq:straina}, \ref{eq:strainb}, and \ref{eq:strainc} in 1 year of integration. Treating LISA as a three interferometer detector, we sum in quadrature the SNRs of the signals produced in the component interferometers to obtain a total SNR. 

\begin{figure}
    \centering
    \includegraphics[width=\columnwidth]{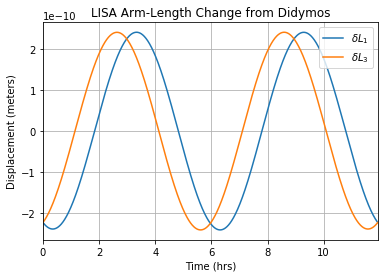}
    \caption{The expected periodic change in length of the two nearest LISA laser arms L1 and L3 as shown in Figure \ref{fig:constellation} produced by the binary asteroid Didymos over one full period of its secondary's orbit. We configure Didymos to be located 450 km away from S/C1 and planar with LISA. The signal is obtained from Equations \ref{eq: displacement} and \ref{eq:length}.}
    \label{fig:Didymos}
\end{figure}

Setting an SNR threshold of 2.5, we find that the maximum distance Didymos could be from LISA and be detected is 450 km. Figure \ref{fig:Didymos} shows the length changes in L1 and L3 produced by Didymos if Didymos were located 450 km from S/C 1 and planar with LISA. Note that the two signals are out of phase due to the different directions of the laser arms. For this contrived configuration, Didymos will produce a signal with a displacement amplitude of $2.4\times10^{-11}$ m in the two nearest laser arms and a negligible displacement in the farther arm. Having an asteroid at a distance of $450$ kilometers from the nearest spacecraft for as long as a year is of course implausible.  Additionally, 450 km is substantially closer than the distance necessary for detecting the already rare case of single asteroid flybys \citep{2006CQGra..23.4939V, 2009CQGra..26h5003T}. It is therefore extremely unlikely that Near-Earth asteroid binaries such as Didymos would be detected by LISA. 

\subsection{Main-Belt Binary Detection}
\label{subsec:mainbeltbin}

\subsubsection{Asteroid Binary Population}
To characterize the set of main belt binary asteroids that can be studied through interferometric detection, we simulate a population of binary asteroids. Since there are an estimated $1-2 \times10^6$ asteroids larger than 1 km in diameter and up to 16\% of asteroids are found in binaries \citep{margot2002binary}, we generate $10^5$ binaries as we expect there to be approximately this many in the belt. We assume an asteroid diameter cumulative distribution of $D_A^{-2.34}$ \citep{2015A&A...578A..42R} and the mass density of asteroids to be 2 g/cm$^3$ \citep{britt2003asteroid, carry2012density}, integrating asteroid diameters from 5 to 1000 km. We assume the asteroids to be spheres and the binaries to have circular Keplerian orbits since most observed binary asteroids have very small eccentricities \citep{MARCHIS200897, 2015aste.book..355M}.  We choose the separation scaling factor $d/D_1$ to follow a Gaussian distribution. For binaries with $D_1 > 20$ km, we give the distribution a mean of 6 and a standard deviation of 2.5 since the majority of observed group L binaries have orbital separations around $6\times D_1$ \citep{2007Icar..190..250P}. For binaries where $D_1<20$ km, we give the distribution a mean of 2.5 and a standard deviation of 1, since the majority of small binaries have separations of $1.25-3.6 D_1$ \citep{2015aste.book..375W}. 

\begin{figure}
    \centering
    \includegraphics[width=\columnwidth]{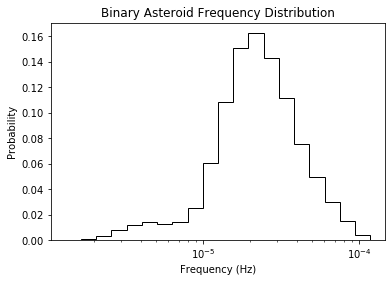}
    \caption{The distribution of signal frequencies expected by binary asteroid systems in the main belt. The signal frequency is twice the orbital frequency.}
\label{fig:AsteroidFreq}
\end{figure}

We plot a histogram of the signal frequencies from the simulated binary asteroids in Figure \ref{fig:AsteroidFreq}. All frequencies are on the order of $10^{-4}$ Hz and below. Approximately {91\%} of our simulated binaries have signal frequencies between $10^{-5}$ and $10^{-4}$ Hz, while {98\%} of these binaries have a frequencies that fall between $4\times10^{-6}$ and $2\times10^{-4}$ Hz. The median frequency of our simulated sample is {$2.2\times10^{-5}$} Hz and the mean frequency is {$2.6\times10^{-5}$} Hz. Only about {0.22\%} of these binaries have frequencies greater than $10^{-4}$ Hz and fall in the LISA sensitivity band shown in Figure \ref{fig:LISAcurve}.

\subsubsection{Binary Asteroid Frequency Disruption}
\label{subsubsec:Binfreqdisrup}
Binary asteroids may be disturbed by close encounters with single asteroids in the asteroid belt. To estimate the probability of such events, we estimate the cross section of disturbing a binary and the flux of asteroids in the belt. We assume that in order for the binary frequency to be noticeably changed, the tidal force on the secondary asteroid in the binary from the single must be greater than $1\%$ of the gravitational force from the primary. This condition can be written as 
\begin{equation}
    4\frac{Md^3}{m_1R^3}>0.01,
\end{equation}
where $m_1$ is the mass of the primary, $d$ is the separation of the masses in the binary, $M$ is the mass of the single asteroid and $R$ is the distance to the single asteroid from the center of the binary.

We estimate the cross sectional area $\sigma$ of a frequency disrupting flyby as 
\begin{equation}
    \sigma=\pi R^2= \pi (\frac{400 M b^3}{m_1})^{\frac{2}{3}}.
\end{equation}
We can use Equations 7 and 8 in \cite{1971NASSP.267..595K} to estimate the number of times a binary is disturbed in 10 years assuming the asteroid spatial density of $7.5\times10^{-19}$ km$^{-3}$ and relative velocity of $4.5$ km/s (See Section \ref{sec:AsteroidBelt}). To obtain an upper limit, we calculate the probability of an encounter for a binary whose primary diameter is $20$ km and separation $d=120$ km and a single asteroid whose diameter is 30 km. The number of frequency disrupting single encounters for this binary is estimated to be $~0.003$ in 10 years. Thus, an upper limit on the percentage of asteroid binaries that will have their frequencies disrupted in a $\sim10$  year observation mission time is $0.3\%$ because this is a particularly large case. Consequently, the overwhelming majority of binary asteroid frequencies should remain stable over the timescales we hope to observe asteroids. 

\subsubsection{Main-Belt Detection with LISA Sensitivity}
We gauge the main-belt binary asteroid detection capabilities of an interferometer with LISA's sensitivity. From the {0.22\%} of asteroid binaries whose signal frequencies fall in the LISA band, we choose the median frequency binary and determine how far away from the detector it could be detected. The binary chosen is a group B binary and has a signal frequency of $1.05\times10^{-4}$ Hz, a primary component with a diameter of {5.5} km, and a secondary component with a diameter of {5.2} km. We configure the binary such that it has angular coordinates $(\theta=0, \phi=0)$ (See Figure \ref{fig:coordinate}) and its orbit is planar with LISA. We calculate the periodic length change in each of the three interferometer arms using Equations \ref{eq: displacement} and \ref{eq:length}. Following the same detection algorithm described in Section \ref{subsec:Didymos}, we calculate the total SNR with LISA sensitivity. 

\begin{figure}
    \centering
    \includegraphics[width=\columnwidth]{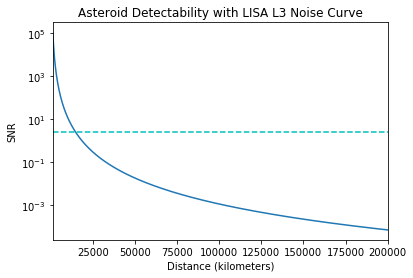}
    \caption{SNR over 1 year of integration time as a function of distance to the nearest LISA S/C for a binary asteroid with component diameters of {5.5} km and {5.2} km (assuming an asteroid density of 2 g/cm$^3$) and a primary signal frequency of $1.05\times10^{-4}$ Hz. The dashed line at an SNR of 2.5 represents a threshold for detection.}
    \label{fig:SNR}
\end{figure}

Figure \ref{fig:SNR} plots the SNR in one year of integration time from the binary asteroid system described above as a function of distance between the center of mass of the binary and the nearest LISA S/C. The binary asteroid can be observed with LISA's sensitivity only in a very small range since the displacement amplitude falls off as $1/r^4$. An interferometer with LISA's sensitivity will only be able to observe small binaries within about 10,000 km of its nearest S/C, a whole two orders of magnitude smaller than LISA's arm-lengths. Furthermore, LISA's orbital radius is 150,000,000 km and would be even larger if it were located in the asteroid belt. Such a long term close encounter between the detector and a binary asteroid would therefore be extremely unlikely to occur. The smallest displacements LISA is capable of detecting within a year, $\sim10^{-13}$ meters, are too limiting for observing binaries at larger distances. 

\subsubsection{Asteroid Belt Binary Signals} 
\label{subsubsec:AsteroidSignals}
To make binary asteroid detections in the asteroid belt, substantially improved interferometer technology is needed. In order to gauge the necessary detection sensitivity, we calculate the signals expected in a space-based interferometer whose three S/Cs have orbital radii at the center of the asteroid belt (i.e.\ 2.7 AU) and are separated by $120^\circ$. This will give the interferometer arm-lengths of $4.6$ AU, allowing the detector to be sensitive to lower frequencies \citep{2016IJMPD..2530001N}. This is the concept for our large AE to be discussed in Section \ref{sec:designandsensitivity}.

\begin{figure}
    \centering
    \includegraphics[width=\columnwidth]{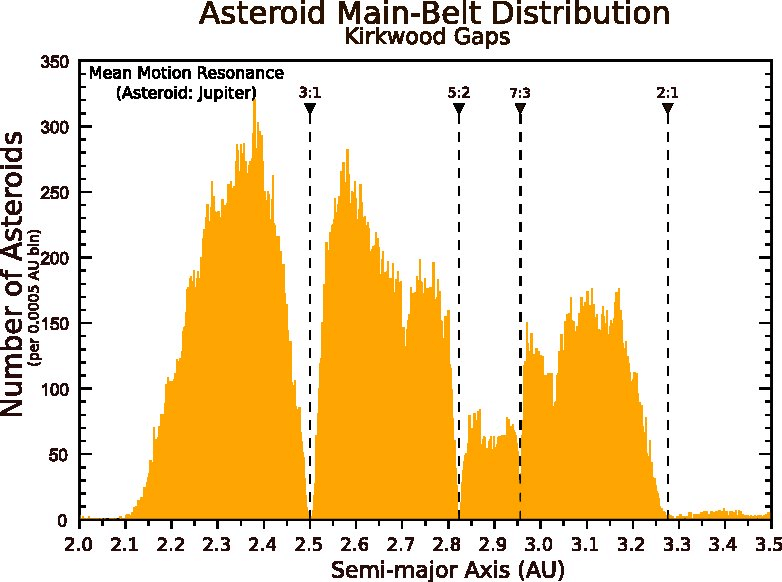}
    \includegraphics[width=\columnwidth]{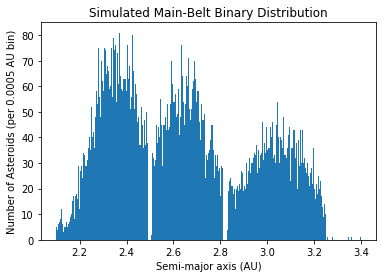}
    \caption{The real semi-major axis distribution of asteroids in the belt including Kirkwood gaps (top) as compared with our simulated semi-major axis distribution of binary asteroids (bottom). We account for the 3:1, 5:2, and 2:1 Kirkwood gaps, neglecting the 7:3 gap as it is not as prominent as the other three. Top panel taken from Figure 2 of \protect\cite{2015arXiv151104835K}.}
    \label{fig:kirkwood}
\end{figure}

We distribute the binary asteroid population across the asteroid belt, qualitatively reproducing the real distribution of asteroids in the belt, and calculate the expected displacement signals in the interferometer. We approximate the radial distribution of asteroids in the asteroid belt by modeling the three regions of the asteroid belt with Gaussian distributions whose means are located at the center of the region. 
For the inner belt, we choose the mean to be 2.35 AU, and the standard deviation to be 0.1 AU. For the middle belt, we choose the mean to be 2.65 AU and the standard deviation to be 0.1 AU while for the outer belt we choose the mean to be 3.05 AU and the standard deviations to be 0.1 AU. We assume the probability of asteroids to be in the inner belt to be 0.4 and the probability of asteroids to be in the middle and outer belt to be 0.3 each. We set the probability of binaries with semi-major axes within 0.01 AU of 2.5 AU and 2.8 AU to be 0 to simulate the prominent 3:1 and 5:2 Kirkwood gaps which contain very few asteroids \citep{1996CeMDA..65..175M}. These choices qualitatively recreate the radial distribution of asteroids in the asteroid belt. Figure \ref{fig:kirkwood} shows the comparison between the observational distribution and our simulated distribution. 

Additionally, the binaries are distributed uniformly 360$^{\circ}$ around the Sun and from $-30^{\circ}$ to $30^{\circ}$ in altitude with respect to the orbital plane. The binaries are given random angular momentum directions with respect to the orbital plane. All signals are calculated assuming the binaries are stationary with respect to the S/Cs.

\begin{figure}
    \centering
    \includegraphics[width=\columnwidth]{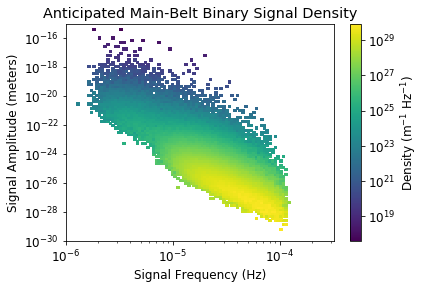}
    \caption{A density plot of the signal amplitudes received by a space-based interferometer and frequencies for the simulated population of $10^5$ binary asteroids distributed throughout the asteroid belt. The signals are calculated assuming a 4.6 AU arm-length detector whose S/Cs are situated inside the asteroid belt in the orbital plane.}
    \label{fig:amplitudes}
\end{figure}

Figure \ref{fig:amplitudes} shows a 2D density plot of the displacement signal amplitudes (summed in quadrature) and the signal frequencies of our simulated binary asteroid population. The amplitudes span 14 orders of magnitude, ranging from $10^{-30}$ m to $10^{-16}$ m. We find that 95\% of simulated binaries have signal amplitudes less than $5\times10^{-22}$ m. The overwhelming majority of these low amplitude binaries have frequencies above $10^{-5}$ Hz. This is because small binaries, which compose 92\% of our binary population, are tightly bound. The higher amplitude and lower frequency binaries are primarily the more loosely bound group L asteroid binaries. All signals are well below LISA sensitivity.

\begin{figure}
    \centering
    \includegraphics[width=\columnwidth]{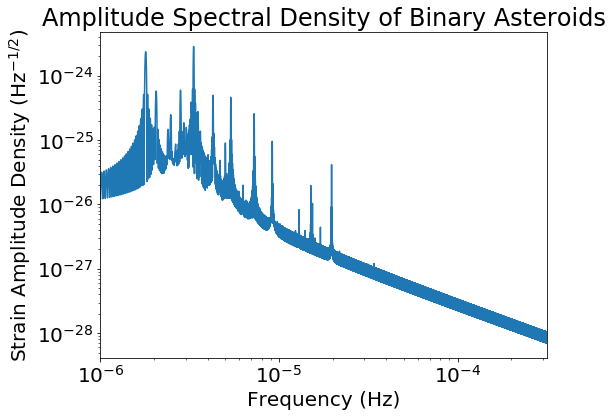}
    \caption{The strain ASD computed for the realization of binary asteroid population used in this paper after 1 year of integration time. The amplitudes used to generate this plot are the same amplitudes computed for Figure \ref{fig:amplitudes}. The ASD peaks at lower frequencies because the largest amplitude binaries have frequencies in the $10^6-10^5$ Hz range, despite there being a higher density of binaries at higher frequencies.}
    \label{fig:ASD_asteroids}
\end{figure}

For LISA, the large number of galactic white dwarf binaries are expected to be a source of confusion noise preventing the frequency resolution of some binaries \citep{2010ApJ...717.1006R}. We expect the problem of distinguishing binary asteroids to be similar to the galactic binary resolution problem. {To determine whether binary asteroids will be resolvable over confusion noise, we compute the power spectral density (PSD) of the total binary asteroid population we created after 1 year of integration time and observe how many individual binary asteroids will have amplitude SNRs higher than 2.5 against this PSD. To compute the PSD for our asteroid population, we individually compute the PSD of each asteroid signal after 1 year of integration time and sum the PSDs together. We take the binary asteroid signals in our detector to be of the form 
\begin{equation}
   s_i(t)= A_i \cos(2\pi f_i t+ \Phi_i)\Theta \left (1- \frac{2|t|}{T} \right ) ,
\end{equation}
where $A_i$ is the strain amplitude of a binary signal (i.e. an amplitude computed for Figure \ref{fig:amplitudes} divided by the interferometer length $L=4.6$ AU), $f_i$ is the signal frequency of a binary, $\Phi_i$ is a phase whose value is random, $\Theta(x)$ is the step function, and $T$ is the integration time of a signal. We include the factor of $\Theta \left (1- \frac{2 |t|}{T} \right ) $ to account for the fact that the signal is only being detected for a finite time $T$. The PSD of a signal is defined as the Fourier transform of the its autocorrelation function. The autocorrelation function of $s(t)$ is 
\begin{equation}
R(\tau)= \left < s(t+\tau)s(t) \right > = \frac{A_i^2}{2}\cos{(2\pi f_i \tau})\left (1-\frac{|\tau|}{T} \right),
\end{equation}
when $|t|\leq T$, and 0 for $|t|\geq T$.
Taking the Fourier transform of the above autocorrelation function, the PSD from one binary as a function of frequency is 
\begin{equation}
    S_{i}(f)=\frac{A_i^2}{4} \left (\frac{\sin(\pi (f-f_i) T)}{\pi (f-f_i)T} \right )^2.
\end{equation}
Summing the individual PSDs from all asteroid binaries, we get a total PSD from the asteroid population
\begin{equation}
    S_{AST}(f)=\sum_{i=1}^ {10^5} S_{i}(f).
\end{equation}
We plot the ASD $S_{AST}^{1/2}(f)$ in Figure \ref{fig:ASD_asteroids}. We compute the SNRs of individual binary asteroid signals against this ASD with their contribution to the total PSD subtracted and determine what percentage of them can be detected with SNRs exceeding 2.5. We find that only $0.02\%$ of our simulated binaries can be detected above this ASD. We create 50 other realizations of our binary asteroid distribution and find that $0.018\%\pm0.006\%$ can be detected above the ASD of the other asteroids. The confusion noise will substantially challenge the resolution of individual binary asteroids.}

{The largest binary asteroids will remain individually resolvable, but it would be extremely difficult to individually resolve smaller binaries with only one detector. If a detector's internal noise PSD can be reduced to below the PSD of the binary asteroids, these binaries asteroids, though unresolved, can still be detected as additional signals in the detector. However, for a binary asteroid mission to be realized and fruitful, a solution to the confusion noise is needed. This challenge can in part be mitigated by the use of matched filtering to detect the largest binaries and the removal of detected signals from the detector data stream \citep[e.g.]{2010PhRvD..81f3008B}. Additionally, the use of multiple constellations to correlate binary signals between multiple detectors can help break the degeneracy. We leave more detailed study of this problem to future work and continue with our analysis assuming that by the time of such an asteroid mission's operation, individual binaries can be detected. 
}

{As noted in Section \ref{subsec:mainbeltbin}, all of our calculations are with circular binaries due to the scarcity of highly eccentric binary asteroids \citep{MARCHIS200897, 2015aste.book..355M}. Eccentric binaries would generate signals with higher harmonics, causing their signal density to bleed further into other frequencies; however, because these eccentric binaries are rare and their signal density is stretched over more frequencies, the signal density of similarly sized circular binary asteroids at a given frequency will exceed that of eccentric binaries. Therefore, the detection of circular binaries should not be affected by the confusion noise of eccentric binary higher harmonics as eccentric binaries are less common and their signals are weaker. The confusion noise of the other circular binaries, due to their large number, will pose a much larger challenge. Nevertheless, eccentric binaries will consequently be more difficult to detect.}

\subsection{Detection of Asteroid Collisions and Close Encounters}
\label{subsec:CollisionsandCloseEncounters}
The detection of asteroid collisions and close encounters in the asteroid belt is of interest to solar system astronomers \citep{1969JGR....74.2531D, 2005Icar..179...63B}. However, signals from asteroid collisions and close encounters are very small. The GW amplitude for Newtonian trajectories is
\begin{equation}
    h_{ij}=\frac{2G}{rc^4}\Ddot{Q}^{TT}_{ij}(t-\frac{r}{c}),
\end{equation}
where $\Ddot{Q}^{TT}_{ij}(t-\frac{r}{c})$ is the second time derivative of the transverse traceless quadrupole moment evaluated at the retarded time, $r$ is the distance to the source, $G$ is the gravitational constant, and $c$ is the speed of light \citep{Flanagan_2005}. For head on collisions between two masses under the influence of gravity, the second time derivative of the quadrupole moment is 
\begin{equation}
    \Ddot{Q}^{TT}_{ij}=\frac{6 GM^2}{b}\begin{pmatrix}
2 & 0 & 0\\
0 & -1 & 0 \\
0 & 0 & -1
\end{pmatrix},
\end{equation}
where $M$ is the sum of the two masses in the system and $b$ is their separation which shrinks over time. Assuming a pair of perfectly spherical asteroids 10 km in diameter with uniform density 2 g/cm$^3$, a minimum separation just before the collision of $b=10$ km, and $r =0.5$ AU, the GW strain is $h\sim\frac{12 G^2 M^2}{rc^4b}=4\times 10^{-38}$. By contrast, for a binary system with the same masses and an orbital separation $d=25$ km, the strain from the Newtonian gravitational force is $h_{\rm Newt}\sim\frac{3G\mu d^2}{ 8\omega^2r_{i}^4L}=4\times10^{-35}$ assuming $L=4.6$ AU. GW signals from such collisions would be much smaller than the Newtonian signals we expect from binaries.

Similarly, GWs from close encounters of asteroids would also be difficult to detect. The GW strain from a parabolic encounter (PE) may be approximated as 
\begin{equation}
   h_{\rm PE} \sim \frac{\sqrt{85}}{4}\frac{G^2}{rc^4}\frac{m_1 m_2}{b},
\end{equation}  where $m_1$ and $m_2$ are the masses of the two asteroids and $b$ is their minimum separation \citep{2006ApJ...648..411K}. For two 10 km diameter masses, $b=25$ km, and $r=0.5$ AU, the PE GW strain is $h\sim7\times10^{-40}$, much smaller than the Newtonian strain expected from binaries. 

In addition, the Newtonian strain from asteroid collisions and parabolic encounters would be very difficult to isolate since it is not periodic. The additional Newtonian force on the detector TM from such an event would likely be indistinguishable from the noise of single asteroids in the vicinity of the detector. Consequently, Newtonian gravitational signals from binary asteroids are the best case for interferometeric asteroid detection.
\section{Asteroid Explorer Design and Sensitivity}
\label{sec:designandsensitivity}

\begin{figure}
    \centering
    \includegraphics[width=\columnwidth]{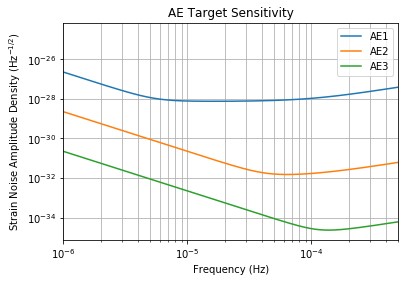}
    \caption{The target strain noise ASD curves for three AE models, space-based interferometers optimized for binary asteroid detection with increasing access to the asteroid belt. In 13.5 years of observation, we expect AE1 to detect $\sim1.5\%$ of binaries, AE2 to detect $\sim21\%$ of binaries, and AE3 to detect $\sim82\%$ of binaries in the belt. These are the same curves for AE shown in Figure \ref{fig:LISAcurve} zoomed in.}
    \label{fig:AEcurve}
\end{figure}

We {envisage} AE as a space-based interferometeric detector utilizing the same principles for observation as the current LISA design. Like LISA, AE will be arranged in an equilateral triangle with its three S/Cs at vertices of the triangle. As detailed in Section \ref{subsec:mainbeltbin} and shown in Figure \ref{fig:amplitudes}, the binary asteroid sources of interest will likely have frequencies in the range $10^{-6}-10^{-4}$ Hz and will induce displacement amplitudes ranging from $10^{-16}-10^{-30}$ meters. This is same the frequency range proposed for $\mu$Ares, whose arm-length may exceed 1 AU \citep{2019arXiv190811391S}. Because of this rather large span of amplitudes, we determine three potential AE sensitivities that have access to increasing percentages of the asteroid belt over a feasible mission lifetime and label them AE1, AE2, and AE3. Figure \ref{fig:AEcurve} shows the three target interferometer sensitivity curves for AE1, AE2, and AE3. For a 13.5 year mission lifetime ($\sim 3$ solar orbital periods of the S/Cs), AE1 will have access to the largest binaries in the belt, detecting 1.5\% of the binary amplitudes shown in Figure \ref{fig:amplitudes} with SNRs larger than 2.5, while AE2 will be able to probe the majority of large asteroid binaries and an increased percentage of smaller asteroids, detecting $21\%$ of amplitudes shown in Figure \ref{fig:amplitudes} with SNRs larger than 2.5. AE3 represents the approximate sensitivity required to thoroughly map the asteroid belt as it is sensitive enough to probe smaller binaries, detecting 82\% of the amplitudes shown in Figure \ref{fig:amplitudes} with SNRs larger than 2.5. 

Parameters such as arm-length, metrology noise, and acceleration noise affect the sensitivity curve of a space-based interferometer \citep{2016IJMPD..2530001N}. \cite{2016IJMPD..2530001N} gives the space-based interferometer ASD as a function of arm-length, metrology noise, and acceleration noise as
\begin{equation}
    S^{1/2}_n(f)= \frac{1}{L}([1+0.5(\frac{f}{f_L})^2]S_p+\frac{4S_a}{{(2\pi f)^4}})^{1/2},
    \label{eq:sensitivity}
\end{equation}
where $L$ is the arm-length of the laser interferometer, $f_L=c/(2\pi L)$ is the arm transfer frequency, $S_p$ is the metrology noise power, and $S_a$ is the acceleration noise power. Table \ref{table:1} shows the arm-length, metrology noise, and acceleration noise requirements for the detectors proposed in this paper. In the following subsections, we discuss in detail the noise floor and orbit requirements of the three AE interferometers.

\begin{table*}
\setlength{\tabcolsep}{16pt}
\begin{tabular}{ p{6cm} c c c  }
 
 \hline\hline
 Detector     & $L$ (km) & $S_p^{1/2}$ (m/Hz$^{1/2}$) & $S_a^{1/2}$ (m s$^{-2}$/Hz$^{1/2}$)\\
 \hline\hline
 LISA & $2.5\times10^{6}$ & $5\times10^{-12}$ & $3\times10^{-15}$ \\
 
 \hline
 AE1 & $6.9\times10^{8}$ &$5\times10^{-17}$& $3 \times 10^{-26}$ \\
 AE2  &$6.9\times10^{8}$ & $8\times10^{-21}$ &$3\times10^{-28}$\\
AE3 & $6.9\times10^{8}$ & $8\times10^{-24}$ & $3\times10^{-30}$\\
\hline
  IABAE1 & $1.5\times10^{8}$ &$5\times10^{-17}$& $3 \times 10^{-26}$ \\
 IABAE2   &$1.5\times10^{8}$ & $8\times10^{-21}$ &$3\times10^{-28}$\\
IABAE3 & $1.5\times10^{8}$ &$8\times10^{-24}$& $3\times10^{-30}$\\
 \hline\hline
\end{tabular}
\caption{The detector specifications we prescribe for the three AE models and the three IABAE models. $L$ is the arm-length, $S_p^{1/2}$ is the metrology noise amplitude, and $S_a^{1/2}$ is the acceleration noise amplitude. We include the specifications of LISA for reference.}
\label{table:1}
\end{table*}

\subsection{Noise Floor Requirements}
\subsubsection{Metrology Noise}
In order to achieve the noise floors shown in Figure \ref{fig:AEcurve}, the shot noise, which is the dominant metrology noise, must be substantially lower than the shot noise in LISA. The expression for laser shot noise in space-based interferometers is
\begin{equation}
    S_p^{1/2}=(\frac{hc\lambda^3L^2}{2 \pi^2 \eta P D_o^4})^{1/2},
    \label{eq:shot}
\end{equation}
where $h$ is Plank's constant, $c$ is the speed of light, $\lambda$ is the wavelength of the laser light, $L$ is the length of the interferometer laser arm, $\eta$ is the optical loss in the optics, $P$ is the laser power, and $D_o$ is the diameter of the collection mirror
\citep{harry2006laser}. The shot noise of LISA will be about $5\times10^{-12}$ m/Hz$^{1/2}$ as LISA will have $L=2.5$ Gm, $D_o=30$ cm, and $\lambda=1064$ nm \citep{2017arXiv170200786A}. The metrology noise of $\mu$Ares is proposed to be $5\times10^{-11}$ m/Hz$^{1/2}$ \citep{2019arXiv190811391S}. DECIGO \citep{kawamura2006japanese, kawamura2011japanese} and BBO \citep{harry2006laser} intend to achieve shot noise amplitude limits of $1\times10^{-17}$ m/Hz$^{1/2}$. 

For AE1, we prescribe a similar metrology noise limit to DECIGO and BBO, $5\times10^{-17}$ m/Hz$^{1/2}$. A metrology noise limit approaching $8\times10^{-21}$ m/Hz$^{1/2}$ is needed to achieve the goals of AE2. AE2 aims to detect displacement amplitudes under $10^{-24}$ m in a decade's worth of observation time. This sensitivity rivals that of proposed next-generation ground-based detectors \citep{2017CQGra..34d4001A, 2019BAAS...51g..35R, Sullivan_2020}, and would be very difficult to achieve in space-based detectors. Since we prescribe AE to have 4.6 AU arm-lengths, the factor $\frac{\lambda^3}{PD_o^{4}}$ in Equation \eqref{eq:shot} for AE2 must be reduced by a factor of $10^{15}$ from LISA. We set the metrology noise limits of AE3 at $8\times10^{-24}$ m/Hz$^{1/2}$, which will require the factor of $\frac{\lambda^3}{PD^{4}}$ be reduced by a factor of $10^{18}$ from LISA.
\subsubsection{Acceleration Noise}
The even greater challenge to achieving the sensitivities shown in Figure \ref{fig:AEcurve}, however, is limiting the acceleration noise. We prescribe AE1 to have $S_a^{1/2}=3 \times 10^{-26}$ m s$^{-2}$/Hz$^{1/2}$, AE2 to have $S_a^{1/2}=3\times10^{-28}$ m s$^{-2}$/Hz$^{1/2}$, and AE3 to have  $S_a^{1/2}=3\times10^{-30}$ m s$^{-2}$/Hz$^{1/2}$. This is $10^{-11}$, $10^{-13}$, and $10^{-15}$ times that of LISA, respectively. This would require substantial technological development in limiting acceleration noise. This advancement must include optimizing power of the laser to lower metrology noise without producing excess acceleration noise \citep{schumaker2003disturbance}. A substantial increase in TM size along with appropriate modifications to the S/C electronics and TM shielding mechanisms would be required. These modifications would need to limit magnetic noise, interference from cosmic rays, and thermal noise \citep{BRAGINSKY20061}.

\subsubsection{Additional Noise Sources}
An additional source of noise arises in the form of the Newtonian gravity gradient from the solar orbit of single asteroids in the asteroid belt \citep{2021PhRvD.103j3017F}. This gradient noise is most prevalent at frequencies between $10^{-9}-10^{-7}$ Hz, but also bleeds into the region between $10^{-6}-10^{-5}$ Hz with strain noise amplitude densities of $10^{-15}$ Hz$^{-1/2}$ and below \citep{2021PhRvD.103j3017F}. The motion of single asteroids thus presents an obstacle to achieving the sensitivities needed for binary asteroid detection. {Mitigating this source of noise is nontrivial, especially since AE would need to be situated inside the asteroid belt to detect binary asteroids. Like for the confusion noise discussed in Section \ref{subsubsec:AsteroidSignals}, one mitigation tactic is to coherently employ multiple detectors and correlate the noise signals \citep{2016PhRvD..93b1101C, 2021PhRvD.103j3017F}. Nevertheless, further study of noise from single asteroids is needed prior to placing AE in the asteroid belt.}

Additionally, AE has a power constraint. The radiation pressure from the Sun will be much lower in the asteroid belt than near the Earth. Consequently, the S/Cs cannot be powered reliably by solar power. Super-ASTROD, a long arm space-based interferometer planned to have S/Cs in orbit at the Lagrange points of Jupiter's orbit, is proposed to use radioisotope thermoelectric generators (RTG) as a power source \citep{2009CQGra..26g5021N}. AE would likely also need to utilize RTGs to serve as its power source. RTGs produce radiation in the form of radioactive decay, which would generate noise effects similar in nature to those of cosmic rays. This would affect detector sensitivity by producing heat and releasing charged particles \citep{BRAGINSKY20061}. The most common radioactive element used in RTGs is Pu-238. Pu-238 radiates alpha particles at a rate of $6.35\times10^{11}$ s$^{-1}$ g$^{-1}$ \citep{LOSALAMOS}, while the cosmic ray flux in the upper atmosphere is on the order of 10$^4$ m$^{-2}$ s$^{-2}$ \citep{stozhkov2001cosmic}. For an RTG composed of 4.5 kg of Pu-238 and an S/C the size of LISA Pathfinder, the radiated charged particles per unit time from the RTG will exceed that of incoming cosmic rays by multiple orders of magnitude. As such, further study into mitigating noise from RTGs would be needed prior to construction to maintain the detector sensitivity while effectively powering the interferometer.

\subsection{Arm-length and Orbital Configuration}
As discussed above, we choose AE's arm-lengths to be 4.6 AU. The three S/Cs will therefore be inside the asteroid belt. The long arm-length serves a dual purpose: 1) making the S/Cs closer to the asteroids and 2) providing sensitivity in the proper frequency range. Longer interferometers are typically more sensitive in lower frequency ranges since they can observe longer wavelength signals \citep{2016IJMPD..2530001N,lisa2000lisa}.  A 4.6 AU arm-length, more than twice that of the proposed ASTROD-GW interferometer \citep{2016IJMPD..2530001N, ni2013astrod, 2010cosp...38.3821N, men2010design}, would give AE a transfer frequency of 93 $\mu$Hz. In contrast, the transfer frequencies of LISA and ASTROD-GW are 19.09 mHz \citep{2019CQGra..36j5011R} and 180 $\mu$Hz \citep{2016IJMPD..2530001N}, respectively.  

We next assess the optimal solar orbit for the S/Cs to maximize detection of the asteroids. To determine the optimal solar orbit for the S/Cs, we use the same simulated population of binary asteroids used to create Figure \ref{fig:amplitudes} and calculate the percentage of detected binaries within a 13.5 year mission time or $\sim 3$ orbital periods for select S/C orbit configurations .  

We examine the detection prospects for the following constellation orbits: 1) a stable uninclined circular orbit whose radii are 2.7 AU (at the center of the asteroid belt), 2) a stable uninclined elliptical orbit whose semi-major axis is 3.1 AU (at the outer edge of the belt) and whose semi-minor axis is 2.3 AU (at the inner edge of the belt), 3) a precessing uninclined elliptical orbit that precesses approximately $10^{\circ}$ per period, and 4) a precessing elliptical orbit inclined $30^{\circ}$ off the elliptic plane. To simulate the motion of the constellation with respect to the asteroid belt, we divide the orbital period into discrete intervals and move the interferometer appropriately in each interval. We calculate the amplitude signal and SNR in each interval and sum in quadrature to obtain a total SNR. For the elliptical orbit, we have the center of the constellation oscillate in one dimension about the Sun with an amplitude of 0.5 AU, with its center shifting after each interval. To simulate the precession, we rotate the LISA triangle 10$^{\circ}$ with respect to the asteroid belt at the end of each full orbital period. Finally, to simulate the inclination of the S/C orbit with respect to the elliptic plane, we oscillate the inclination angle of the whole LISA constellation with respect to the orbital plane between $-30^{\circ}$ and $30^{\circ}$ over the course of one orbital period, changing the inclination after each interval. We keep the binaries stationary with respect to the asteroid belt and only change their position relative to the AE constellation after 1 interval. We choose an SNR detection threshold of 2.5.

We calculate the percentage of binaries throughout the entire belt that can be detected. We find the following results for AE1: 1) detects 1.5\% of the generated binaries; 2) detects 1.7\% of the generated binaries; 3) detects 2.4\% of binaries; and 4) detects 3.6\% of the binaries. For AE2 we find the following: 1) detects 20.9\% of binaries, 2) detects 21.6\% of binaries, 3) 25.7\% of binaries, and 4) detects 35.8\% of binaries. For AE3 we find these results: 1) detects 82.0\% of binaries, 2) detects 82.2\% of binaries, 3) detects 85.9\% of binaries, and 4) detects 89.6\% of binaries.
\begin{figure}
    \centering
    \includegraphics[width=\columnwidth]{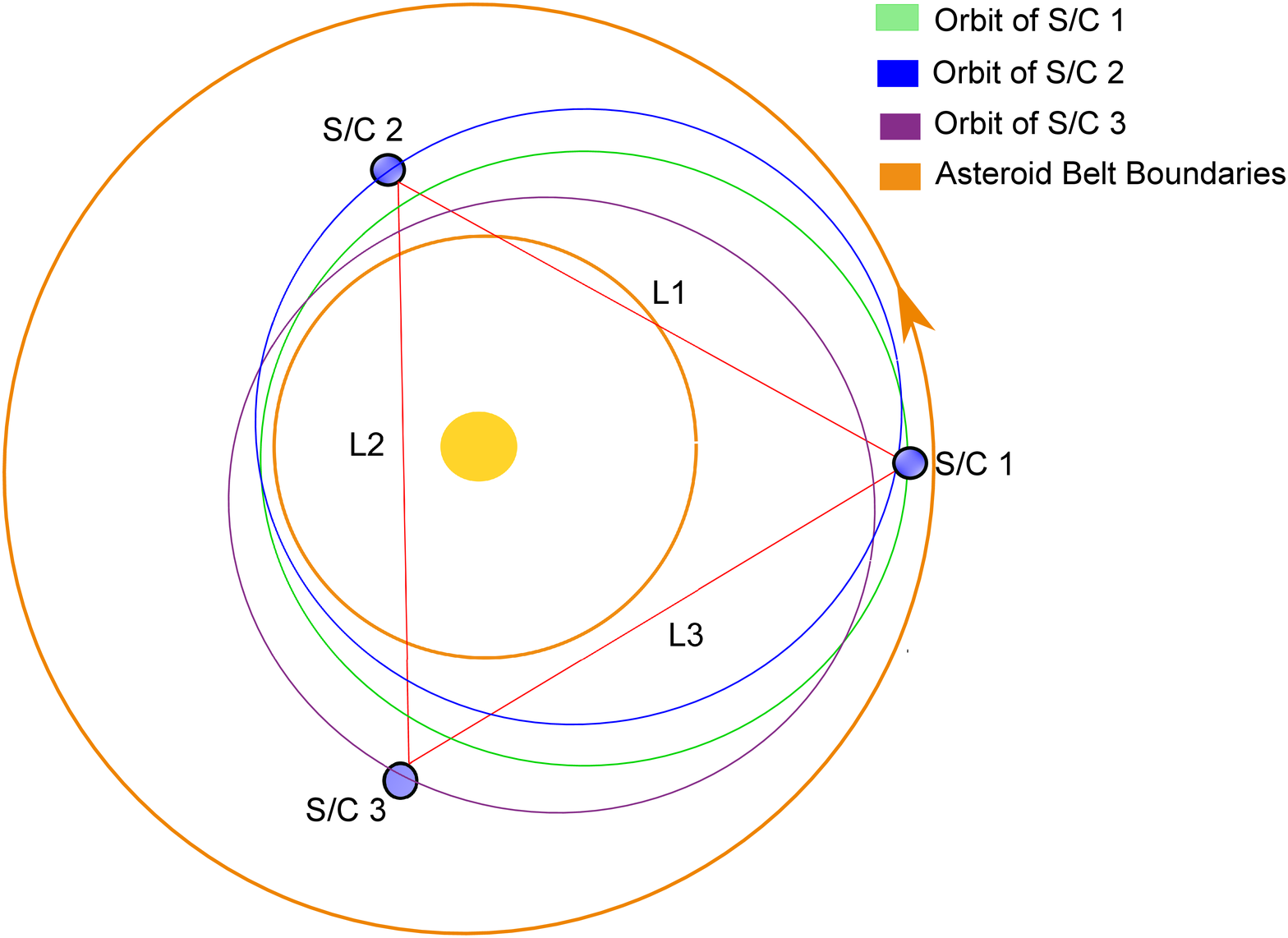}
    \includegraphics[width=\columnwidth]{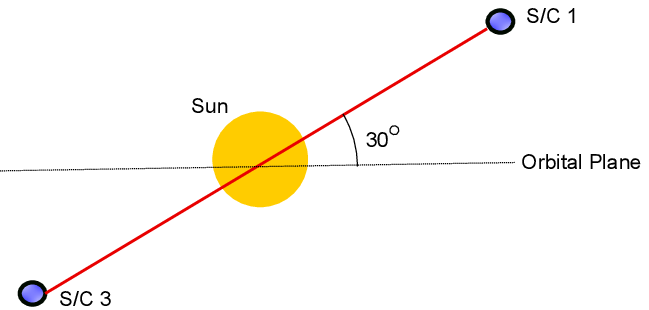}
    \caption{(Top) A top view of the AE configuration and elliptical orbits of the three AE S/Cs in the asteroid belt. The arrow on the boundary denotes the direction the asteroids move around the sun. (Bottom) A side view of the configuration of AE showing the 30 degree inclination of the S/C orbits with respect to the orbital plane.}
    \label{fig:Orbits}
\end{figure}

These results suggest that the number of detected events does not depend as strongly on the orbit of the S/Cs as on sensitivity. Nevertheless, the results for AE1 and AE2 indicate that the use of an inclined elliptical orbit with precession noticeably improves the number of binaries we can expect to detect. Consequently, we select orbit 4) which maximizes the amount of binaries that can be detected and minimizes the area of the belt through which the laser must pass at any given time. Figure \ref{fig:Orbits} shows a top and side view of this orbit. More detailed study will be needed to determine the precise orbit the S/Cs must follow and how to protect the laser arms from potential dust in the asteroid belt.

\subsection{Angular Localization}
\label{subsec:Angloc1}
AE must have source localization capabilities to successfully map the asteroid belt. Earth-based GW detector networks require two or more detectors at different points on the Earth to localize GW sources. Having a network of interferometers at multiple locations allows for triangulation of the GW signal not possible with one stationary interferometer. In contrast, LISA can independently localize a GW source because LISA's position relative to the source will change over the course of its orbit. Because the length of the laser arms of AE is of a similar order as the distances to the asteroid binaries, AE will not be able to localize via parallax. Instead, the different responses of the three AE component interferometers may be used to localize the detected asteroid binaries.  
 
\begin{figure*}
    \centering
    \subfigure[]{\includegraphics[width=0.88\linewidth]{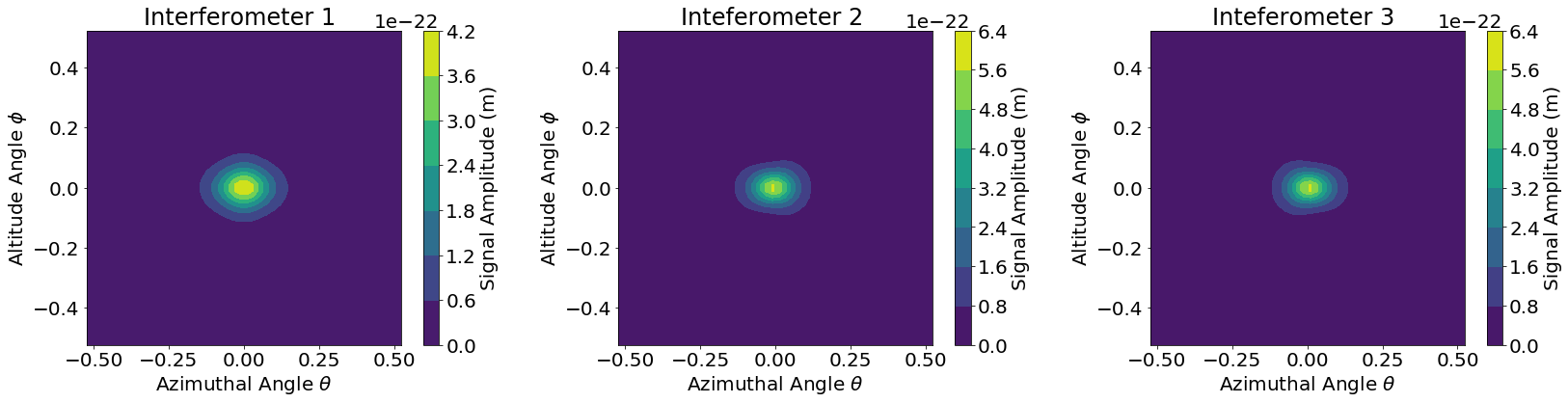}}
    \subfigure[]{\includegraphics[width=0.88\linewidth]{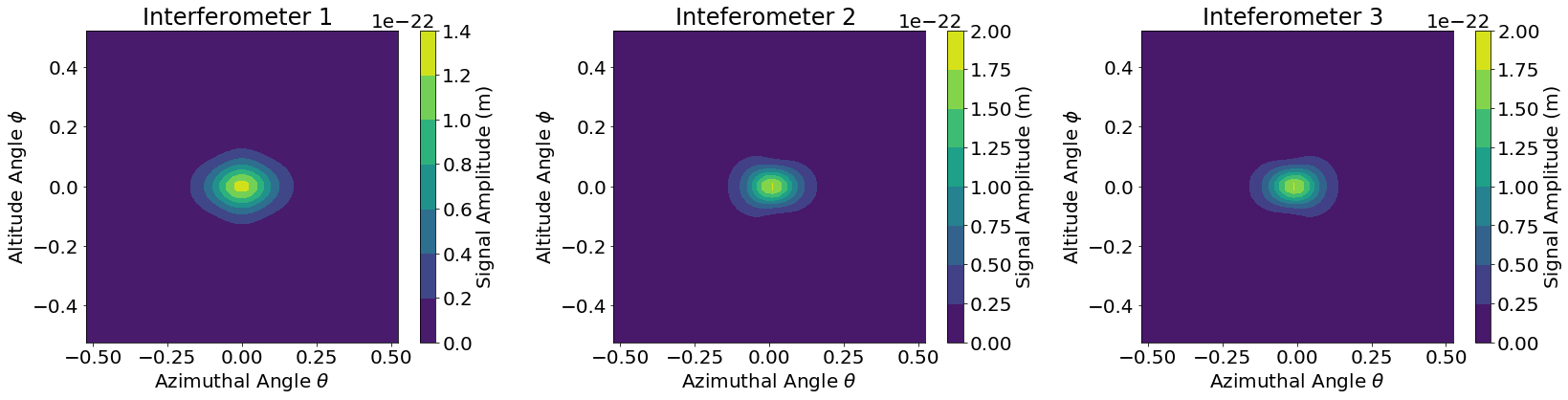}}
    
    \caption{Contour plots of displacement signal amplitude from a binary whose asteroids are each 10 km in diameter as a function of angular position (in radians) in the three component interferometers. Interferometer 1 is the interferometer whose corner is located at azimuthal angle $\theta=0$, Interferometer 2 is the interferometer whose corner is at $\theta=\frac{2\pi}{3}$, and Interferometer 3 is interferometer whose corner is at $\theta=\frac{4\pi}{3}$. The binary is 2.4 AU from the center of AE in (a) while the binary is 3.0 AU from the center of AE in (b). }
    \label{fig:contouroutside}
\end{figure*}

To estimate the localization capabilities, we determine whether the signal from a binary at a specific angular location in the sky can be distinguished from a signal of the same binary at other angular locations. We fix the asteroid diameters and separation as well as the distance from the binary to the center of the AE constellation. We set the plane of the binary to be parallel to the plane of the detector. We calculate the signal from this binary at azimuthal angles $\theta\in [-\frac{\pi}{3}, \frac{\pi}{3}]$ and at altitude angles $\phi \in [-\frac{\pi}{5}, \frac{\pi}{5}]$. We place these limits on the altitude angle because asteroids in the belt have inclinations to the orbital plane of generally no more than 30 degrees. The point $(\theta, \phi)=(0,0)$ corresponds to the angular location of the nearest AE spacecraft (see Figure \ref{fig:coordinate}).  We compute the signals in the three component interferometers of AE, labeling them Interferometers 1, 2, and 3 (not to be confused with the three AE models AE1, AE2, and AE3). Interferometer 1 is the interferometer whose corner is located at azimuthal angle $\theta=0$, Interferometer 2 is the interferometer whose corner is at $\theta=\frac{2\pi}{3}$, and Interferometer 3 is interferometer whose corner is at $\theta=\frac{4\pi}{3}$. Because AE is an equilateral triangle, only the ordering of component interferometers in the cases with $\theta \in [\frac{\pi}{3}, \pi]$ and $\theta \in [\pi, \frac{5\pi}{3}]$ will differ from the case presented. We perform these calculations at two distances from the center of the AE constellation, one inside the constellation at 2.4 AU and one outside the constellation a distance of 3.0 AU. We also use two different binary sizes: one small binary whose primary and secondary are each 10 km in diameter and whose separation is 25 km, and one large binary whose primary and secondary are each 30 km in diameter and whose separation is 180 km.  Figure \ref{fig:contouroutside} shows contour plots of the signal amplitudes in the three component interferometers as a function of angular position for the small binary. 

We divide the sky region defined by $\theta \in [-\frac{\pi}{3}, \frac{\pi}{3}]$ and $\phi \in [-\frac{\pi}{5}, \frac{\pi}{5}]$  into 2500 pixels equally-spaced in $\theta$ and $\phi$ and estimate how precisely a binary asteroid event located at each of these pixels can be localized for the three AE models. The criterion used to determine whether an event whose angular position is $(\theta_1, \phi_1)$ can be distinguished from an event whose angular position is $(\theta_2, \phi_2)$ is
\begin{equation}
\label{eq:localcond}
    \frac{|h(\theta_1, \phi_1)-h(\theta_2, \phi_2)|}{h{(\theta_1, \phi_1)}}>\rho(\theta_1, \phi_1)^{-1},
\end{equation}
where $h(\theta, \phi)$ is the signal amplitude at a given angular location $(\theta, \phi)$ and $\rho(\theta, \phi)$ is the SNR at a given location $(\theta, \phi)$.
If the condition in Equation \ref{eq:localcond} is not met, we consider the event whose angular position is $(\theta_1, \phi_1)$ to be indistinguishable from an event located at $(\theta_2, \phi_2)$. 

\begin{figure*}
    \centering
    \subfigure[]{\includegraphics[width=\linewidth]{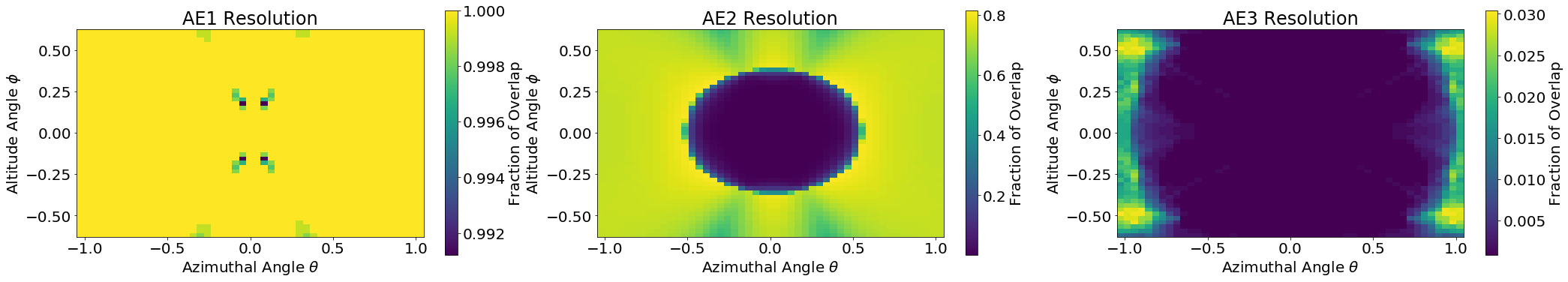}}
    \subfigure[]{\includegraphics[width=\linewidth]{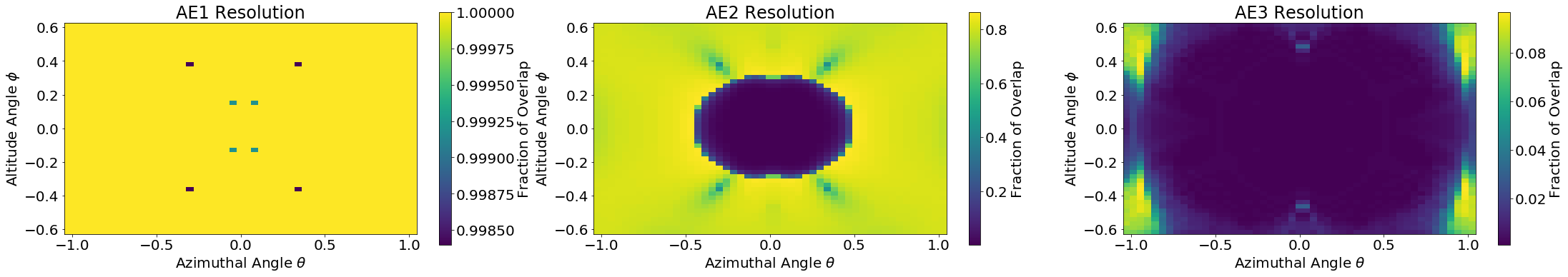}}
    \subfigure[]{\includegraphics[width=\linewidth]{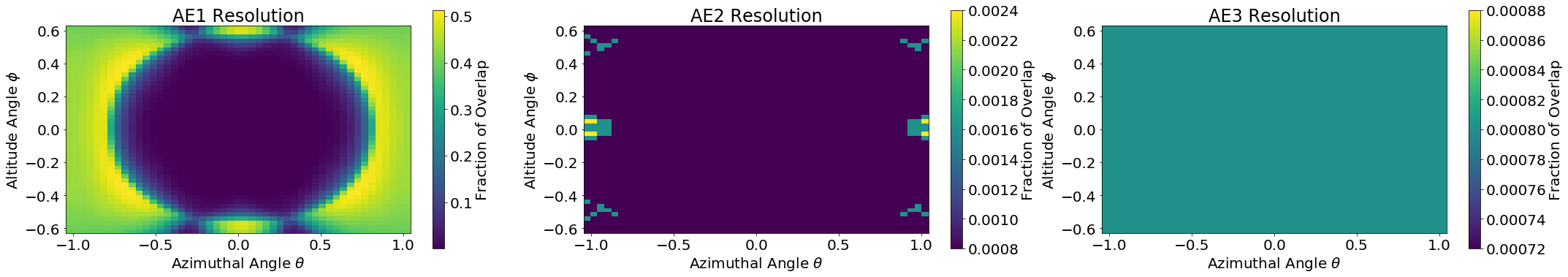}}
    \subfigure[]{\includegraphics[width=\linewidth]{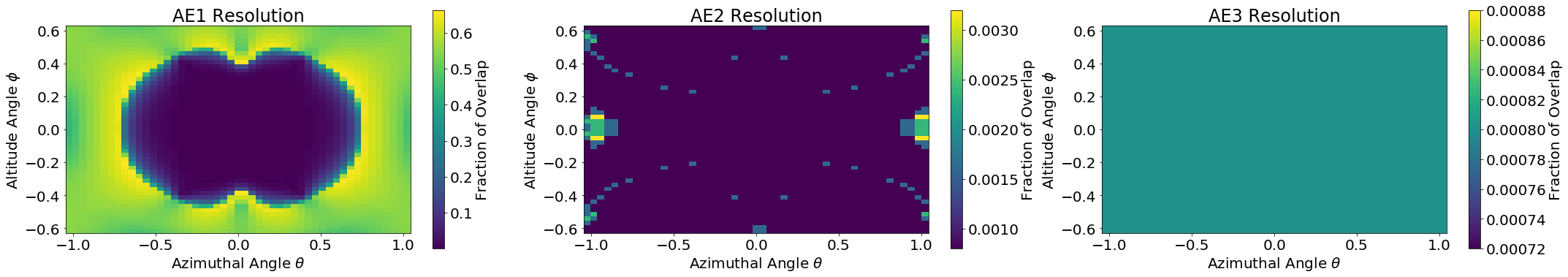}}
    \caption{Plots of the fraction of pixels in the sky region sampled that are indistinguishable from an asteroid binary at given point after 5 years of observation. Plots are shown for each of the three AE models. (a) A binary whose component diameters are 10 km and asteroid separation is 25 km at 2.4 AU from the center of the AE constellation. (b) A binary whose component diameters are 10 km and asteroid separation is 25 km at 3.0 AU from the center of the AE constellation. (c) A binary whose component diameters are 30 km and asteroid separation is 180 km at 2.4 AU from the center of the AE constellation. (d) A binary whose component diameters are 30 km and asteroid separation is 180 km at 3.0 AU from the center of the AE constellation.}
    \label{fig:Resolution}
\end{figure*}

Figure \ref{fig:Resolution} shows plots of the fraction the sky region indistinguishable from an event at a given pixel after 5 years of observation with the three AE models for the 10 km binary and 30 km binary at 2.4 AU and 3.0 AU from the center of the constellation. From these plots, we may estimate the precision of the angular localization abilities of the three AE models for various binary sizes. 

All three models will have some difficulty localizing small binaries. AE1 has virtually no ability to localize these binaries since an event located at every pixel is indistinguishable from at least 99\% of the remaining region. For AE2, events at 15\% of pixels in the sky region sampled are distinguishable from more than 90\% of the remaining sky region for binaries located at both 2.4 AU and 3.0 AU. For AE3, events from all pixels are distinguishable from more than 90\% of the remaining sky region for both 2.4 AU and 3.0 AU.

For 30 km binary asteroids, the localization ability of the three models is much improved. For AE1, 46\% and 33\% of locations in the sky region sampled are distinguishable from more than 90\% of the remaining sky region for 2.4 AU and 3.0 AU, respectively. AE2 and AE3 can localize events located at all pixels in the region with even better precision. AE2 can distinguish events at all pixels from more than 99.65\% of the remaining sky region at both 2.4 AU and 3.0 AU. AE3 can distinguish events at all pixels from more than 99.9\% of the remaining sky region. The precision in the localization for AE2 and AE3 is primarily limited by the symmetry of the detector about the $\phi=0$ plane which causes two events located at angular positions $(\theta, \phi)$ and $(\theta, -\phi)$ to be indistinguishable. 

The assessment of angular localization capabilities reported here is a conservative estimate as the relative delay in response time of the three interferometers and motion of the detector relative to the sources may also be used to further localize detected binary asteroids.

\section{Inner Asteroid Belt Asteroid Explorer}
\label{sec:InnerAsteroidbelt}
\begin{figure}
    \centering
    \includegraphics[width=\columnwidth]{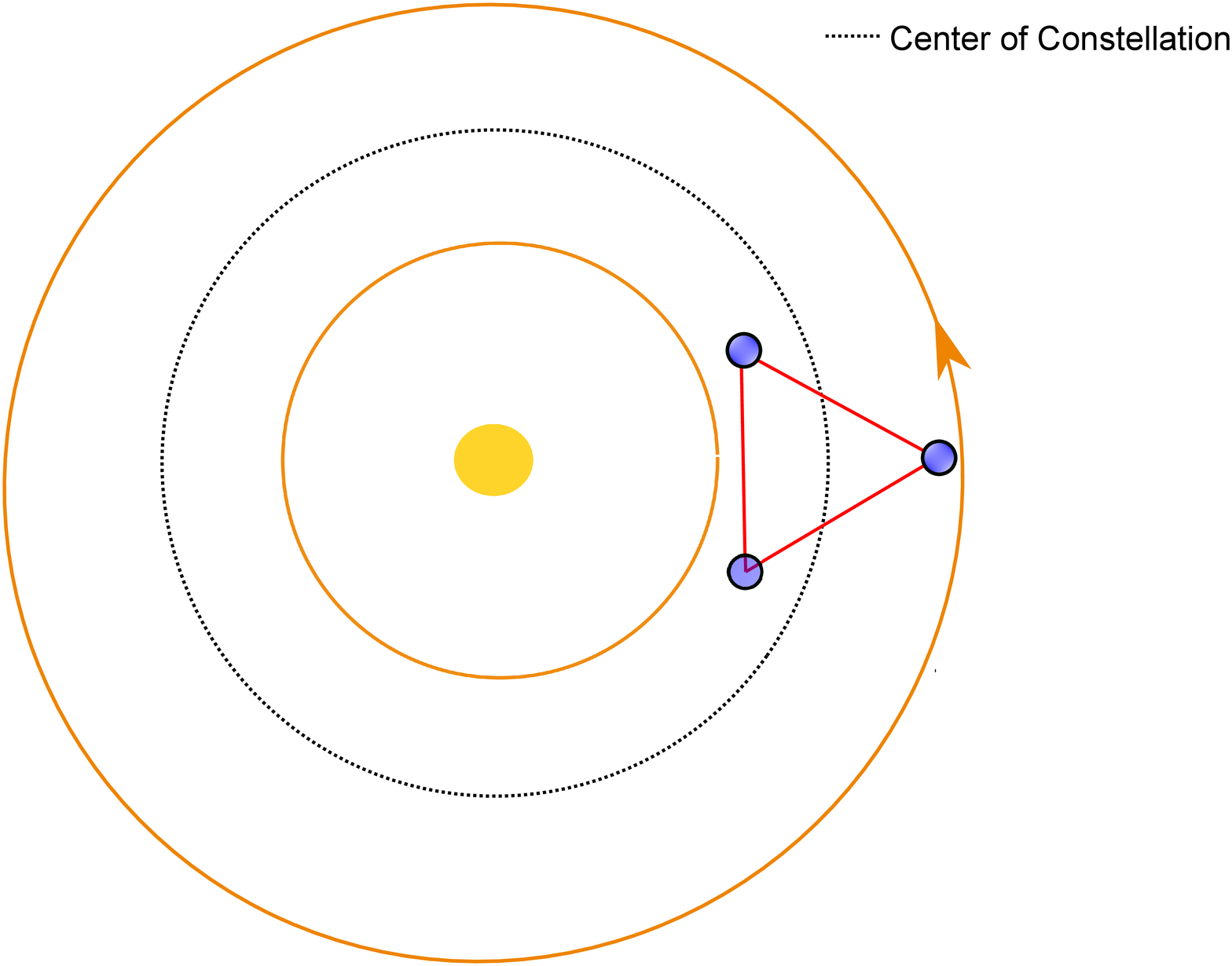}
   
    \caption{A diagram showing the placement of IABAE in the asteroid belt including a dashed line representing that the center of IABAE would precess through the belt over its mission lifetime. The arrow on the boundary denotes the direction the asteroids move around the sun.}
    \label{fig:SmallOrbits}
\end{figure}

We consider the performance of IABAE, a space-based asteroid explorer whose orbit would be confined entirely to the asteroid belt. IABAE would have laser arms with a length of approximately 1 AU and have its center located in the middle of the asteroid belt, 2.7 AU from the sun. IABAE's orbit would precess around the asteroid belt so that it could be sensitive to asteroid binaries at different locations in the belt. Figure \ref{fig:SmallOrbits} shows the placement and orbit of IABAE in the asteroid belt. 

\begin{figure}
    \centering
    \includegraphics[width=\columnwidth]{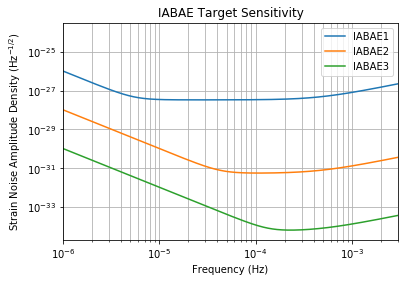}
    \caption{The target sensitivity we propose for the three IABAE models. These detectors have the same metrology and acceleration noise requirements as the three AE models but have 1 AU laser arms rather than 4.6 AU arms. These are the same curves for IABAE shown in Figure \ref{fig:LISAcurve} zoomed in.}
    \label{fig:IABAESensitivity}
\end{figure}

We propose three IABAE models: IABAE1, IABAE2, and IABAE3. We prescribe the same metrology and acceleration noises prescribed to the three AE models AE1, AE2, and AE3. Figure \ref{fig:IABAESensitivity} shows the target sensitivity curves of the three IABAE models. We note that because of the smaller arm-lengths of IABAE, these noise floors will be more achievable than the noise floors of AE.
\subsection{Detection Performance}
We probe the performance of IABAE by calculating the number of binaries that the three IABAE sensitivities shown in Figure \ref{fig:IABAESensitivity} can detect in the mission lifetime. Because IABAE is smaller than AE, it will require a much longer mission lifetime to detect asteroid binaries on all sides of the belt. We consider the orbit of IABAE to precess $10^\circ$ at the completion of one orbital period of the detector, approximately every 4.5 years. Consequently, to reach the whole belt, one would need $\sim160$ years of observation or potentially only 4.5 years of observation with 36 detectors whose centers are spaced $10^\circ$ apart in the orbital plane. We note that because in every orbital period, IABAE will be detecting different sets of asteroid binaries, the probability of binary asteroid frequencies being disrupted during observation remains low (see Section \ref{subsubsec:Binfreqdisrup}). In 4.5 years of observation or the equivalent of 1 orbital period, we find that IABAE1 can detect $1.3\%$ of our simulated binary population, IABAE2 can detect $5.5\%$, and IABAE3 can detect $56.1\%$ with SNRs above 2.5. This performance is not dramatically lower than that of the larger AE model, thus suggesting that even one cycle of observation with this smaller detector could yield fruitful results. After 36 orbital periods, we find that IABAE1 can detect {6.2\%} of our binaries, IABAE2 can detect {56.9\%} of our binaries, and IABAE3 can detect 99.9\% of our binaries. 
\subsection{Angular Localization}
\begin{figure*}
    \centering
    \subfigure[]{\includegraphics[width=\linewidth]{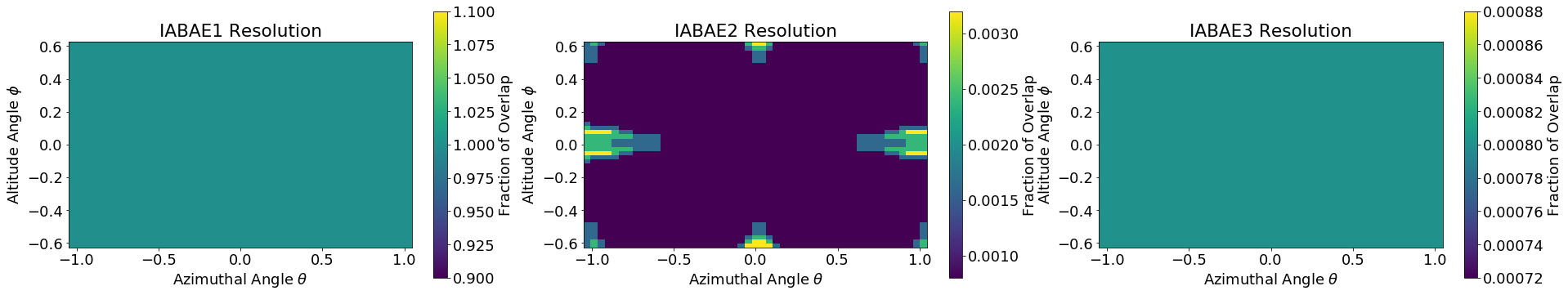}}
    \subfigure[]{\includegraphics[width=\linewidth]{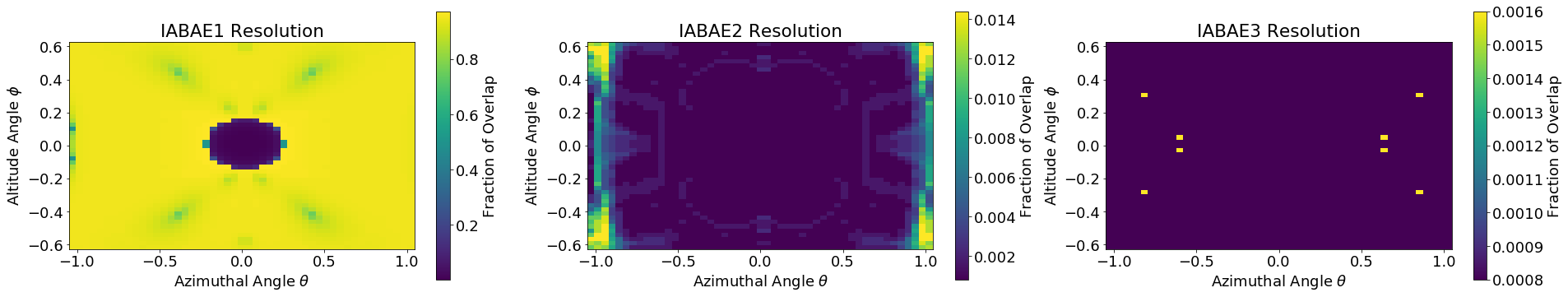}}

    \caption{Plots of the fraction of pixels in the region sampled that are indistinguishable from a given pixel as observed by each of the three IABAE models. (a) A binary whose component diameters are 10 km and asteroid separation is 25 km at 0.3 AU from the center of the IABAE constellation. (b) A binary whose component diameters are 10 km and asteroid separation is 25 km at 0.7 AU from the center of the IABAE constellation.}
    \label{fig:Resolution2}
\end{figure*}

We probe the angular localization capabilities of IABAE in a fashion identical to that described in Section \ref{subsec:Angloc1}. We consider the ability of the three IABAE sensitivities to localize asteroid binaries in the sky region $\theta \in [-\frac{\pi}{3},\frac{\pi}{3}]$ and $\phi \in [-\frac{\pi}{5}, \frac{\pi}{5}]$ and divide the region into 2500 pixels. Because IABAE will be confined to one portion of the belt, we only consider its ability to localize asteroids in its immediate vicinity, within 0.7 AU of its center. We study the localization capabilities of a small binary whose component diameters are each 10 km and whose separation is 25 km. We study its localization capabilities at two distances: 0.3 AU and 0.7 AU. Figure \ref{fig:Resolution2} shows the fraction of the sky region indistinguishable from an event located at a given pixel after 5 years of observation with the three IABAE models. 

IABAE1 will have no ability to localize small binaries located inside the constellation; however, IABAE2 will be able to distinguish events from all locations in the sky region considered from 99.5\% of the rest of the sky region and IABAE3 will be able to distinguish events located at all pixels in the sky region from 99.92\% of the sky region. Outside the constellation, IABAE1 will be able to distinguish 3.7\% of the locations from 90\% of the remaining sky region, while events from the vast majority $(\sim90\%)$ of angular positions will be practically indistinguishable from the rest of the region. IABAE2 will be able to distinguish events located at all pixels from at least 98\% of the remaining sky region with the vast majority of locations distinguishable from more than 99.5\% of the remaining region. IABAE3 will be able to distinguish events located at all pixels from more than 98\% of the sky region. Events from 99.7\% of angular positions will be distinguishable from more than 99.9\% of the sky region sampled with IABAE3. 

For the large binary considered in Section \ref{subsec:Angloc1}, each of the three IABAE models will be able to distinguish events from all pixels from more than 99\% of the remaining sky region. 

\section{Sensitivity to Other Gravitational Wave Sources}
\label{sec:OtherGWs}
With their vastly increased sensitivity in the $10^{-6}$-$10^{-4}$ Hz range, both AE and IABAE would be sensitive to a number of additional astrophysical GW sources. With the same frequency range as $\mu$Ares \citep{2019arXiv190811391S}, we expect AE and IABAE to receive GWs from numerous super-massive black hole (SMBH) binary inspirals, extreme mass ratio inspirals (EMRIs), galactic white dwarf binaries, and potentially the stochastic background. These are the standard anticipated sources in the low frequency GW regime \citep{Biesiada:2018nki}.

Additionally, astrophysical GWs from compact binaries will be detectable with enormous SNRs. AE and IABAE could detect neutron star binaries with component masses of 1.4 M$_\odot$ that are $10^{13}$ years away from merging out to 10 Gpc with SNRs of $\sim100$ in 1 year of observation. Observation of such systems far from merger would directly probe mechanisms that cause compact objects to merge \citep[e.g.]{2021hgwa.bookE...4M}. 
Below $10^{-2}$ Hz, galactic binaries including white dwarf binaries are expected to be present in the LISA and $\mu$Ares band \citep{2010ApJ...717.1006R,  2017arXiv170200786A, 2019arXiv190811391S, 2020A&A...638A.153K}. A unique source of gravitational waves, the quasi-normal modes of SMBHs at very far distances, may also be detectable with potentially very high SNRs \citep{2009LRR....12....2S}. The extra-galactic background of binary sources are expected to have strain amplitudes of $10^{-19}-10^{-17}$ Hz$^{-1/2}$ between frequencies of $5\times10^{-6}$ and $5\times10^{-5}$ Hz with smaller amplitude background sources at frequencies above $5\times10^{-5}$  Hz \citep{2001MNRAS.324..797S}. These sources may contribute to a very large sum of confusion noise in AE and IABAE.  {While GWs will constitute noise when searching for asteroid sources, the sensitivities prescribed for AE and IABAE can detect these sources and allow for their signal to be subtracted from the detectors' data streams. As with confusion noise mitigation, signal processing techniques such as matched filtering can allow for the removal of previously detected signals \citep[e.g.]{2010PhRvD..81f3008B}. Additionally, binary asteroid sources can be distinguished from other astrophysical binary sources because asteroids will be moving relative to the detector. Nevertheless,} further research and development will be needed to isolate the detectors from GW background sources and ensure a successful asteroid detection mission.   
 
AE and IABAE would contribute to the network of interferometric detectors and could serve as a useful tool for astrophysical GW detection with the sensitivity prescribed.  If the sensitivity prescribed is realizable, AE or IABAE could eventually become a successor to the $\mu$Ares detector in the search for GWs in the $\mu$Hz frequency range in addition to an asteroid detection mission. 
\section{Conclusion}
\label{sec:conc}
We have presented a preliminary study into the detection of binary asteroids by interferometric astronomical techniques. We have shown that it is possible to detect a binary asteroid system at very short distances using a space-based GW interferometer. LISA will only be capable of detecting Near-Earth asteroid binaries the size of Didymos $450$ km away from its nearest S/C and main belt-sized binary asteroids with frequencies greater than $10^{-4}$ Hz within a couple 10,000 km of its nearest S/C. Since this distance and frequency range is not feasible for asteroid belt studies, we have proposed AE and IABAE, interferometers configured for detecting binary asteroids.

Our three potential AE models each have 4.6 AU arm-lengths with varied sensitivity prescriptions. They have access to an increased percentage of the asteroid binaries located throughout the asteroid belt. The sensitivities we propose, AE1, AE2, and AE3, can detect as much as $3.6\%$, $35.6\%$, and $89.6\%$ of the binaries in the asteroid belt, respectively. Our IABAE models have arm-lengths of 1 AU with varying sensitivities. IABAE1, IABAE2, and IABAE3 can detect as much as 6.2\%, 56.9\%, and 99.9\% of binaries in IABAE's much longer mission time.

Substantial improvements in technology and background noise removal techniques must be made in order to realize the sensitivities that are necessary for AE and IABAE. These devices will need to be at least $10^{11}$ times more sensitive than LISA in a lower frequency range. 
Nevertheless, if this sensitivity can be realized, a mission dedicated to studying gravitational effects of asteroids in addition to already existent observational surveys will allow for detailed studies of asteroid structure and density. With a dedicated mission for studying asteroids via interferometry, new insight into the evolution of the solar system can be attained.  
\section*{Acknowledgements}
The authors thank Columbia University in the City of New York and the University of Florida for their generous support.
The Columbia Experimental Gravity group is grateful for the generous support of Columbia University. The authors thank Charles Hailey for valuable discussions about the manuscript. AS is grateful for the support of the Columbia College Science Research Fellows program and the Heinrich, CC Summer Research Fellowship. DV is grateful to the Jacob Shaham Fellowship. IB acknowledges support from the Alfred P. Sloan Foundation and from the National Science Foundation under grant numbers 1911796 and 2110060.

\section*{Data Availability}
The data underlying this article will be shared on reasonable request to the corresponding author.

\bibliography{Refs}

\begin{thebibliography}{}
\makeatletter
\relax
\def\mn@urlcharsother{\let\do\@makeother \do\$\do\&\do\#\do\^\do\_\do\%\do\~}
\def\mn@doi{\begingroup\mn@urlcharsother \@ifnextchar [ {\mn@doi@}
  {\mn@doi@[]}}
\def\mn@doi@[#1]#2{\def\@tempa{#1}\ifx\@tempa\@empty \href
  {http://dx.doi.org/#2} {doi:#2}\else \href {http://dx.doi.org/#2} {#1}\fi
  \endgroup}
\def\mn@eprint#1#2{\mn@eprint@#1:#2::\@nil}
\def\mn@eprint@arXiv#1{\href {http://arxiv.org/abs/#1} {{\tt arXiv:#1}}}
\def\mn@eprint@dblp#1{\href {http://dblp.uni-trier.de/rec/bibtex/#1.xml}
  {dblp:#1}}
\def\mn@eprint@#1:#2:#3:#4\@nil{\def\@tempa {#1}\def\@tempb {#2}\def\@tempc
  {#3}\ifx \@tempc \@empty \let \@tempc \@tempb \let \@tempb \@tempa \fi \ifx
  \@tempb \@empty \def\@tempb {arXiv}\fi \@ifundefined
  {mn@eprint@\@tempb}{\@tempb:\@tempc}{\expandafter \expandafter \csname
  mn@eprint@\@tempb\endcsname \expandafter{\@tempc}}}

\bibitem[\protect\citeauthoryear{Abbott et~al.,}{Abbott
  et~al.}{2016}]{abbott2016observation}
Abbott B.~P.,  et~al., 2016, \prl, 116, 061102

\bibitem[\protect\citeauthoryear{{Abbott} et~al.,}{{Abbott}
  et~al.}{2017a}]{2017CQGra..34d4001A}
{Abbott} B.~P.,  et~al., 2017a, \mn@doi [Classical and Quantum Gravity]
  {10.1088/1361-6382/aa51f4}, \href
  {https://ui.adsabs.harvard.edu/abs/2017CQGra..34d4001A} {34, 044001}

\bibitem[\protect\citeauthoryear{Abbott et~al.,}{Abbott
  et~al.}{2017b}]{abbott2017gw170817}
Abbott B.~P.,  et~al., 2017b, \prl, 119, 161101

\bibitem[\protect\citeauthoryear{Abbott et~al.,}{Abbott
  et~al.}{2019}]{abbott2019gwtc}
Abbott B.,  et~al., 2019, Physical Review X, 9, 031040

\bibitem[\protect\citeauthoryear{{Abbott} et~al.}{{Abbott}
  et~al.}{2021}]{2021arXiv211103606T}
{Abbott} R.,  et~al., 2021, arXiv e-prints, \href
  {https://ui.adsabs.harvard.edu/abs/2021arXiv211103606T} {p. arXiv:2111.03606}

\bibitem[\protect\citeauthoryear{{Agrusa} et~al.,}{{Agrusa}
  et~al.}{2020}]{2020Icar..34913849A}
{Agrusa} H.~F.,  et~al., 2020, \mn@doi [\icarus]
  {10.1016/j.icarus.2020.113849}, \href
  {https://ui.adsabs.harvard.edu/abs/2020Icar..34913849A} {349, 113849}

\bibitem[\protect\citeauthoryear{{Amaro-Seoane} et~al.,}{{Amaro-Seoane}
  et~al.}{2017}]{2017arXiv170200786A}
{Amaro-Seoane} P.,  et~al., 2017, arXiv e-prints, \href
  {https://ui.adsabs.harvard.edu/abs/2017arXiv170200786A} {p. arXiv:1702.00786}

\bibitem[\protect\citeauthoryear{{Armano} et~al.,}{{Armano}
  et~al.}{2016}]{2016PhRvL.116w1101A}
{Armano} M.,  et~al., 2016, \mn@doi [\prl] {10.1103/PhysRevLett.116.231101},
  \href {https://ui.adsabs.harvard.edu/abs/2016PhRvL.116w1101A} {116, 231101}

\bibitem[\protect\citeauthoryear{Armano et~al.,}{Armano
  et~al.}{2018}]{PhysRevLett.120.061101}
Armano M.,  et~al., 2018, \mn@doi [Phys. Rev. Lett.]
  {10.1103/PhysRevLett.120.061101}, 120, 061101

\bibitem[\protect\citeauthoryear{{Armstrong}, {Estabrook}  \&
  {Tinto}}{{Armstrong} et~al.}{1999}]{1999ApJ...527..814A}
{Armstrong} J.~W.,  {Estabrook} F.~B.,   {Tinto} M.,  1999, \mn@doi [\apj]
  {10.1086/308110}, \href
  {https://ui.adsabs.harvard.edu/abs/1999ApJ...527..814A} {527, 814}

\bibitem[\protect\citeauthoryear{{Bailes} et~al.}{{Bailes}
  et~al.}{2021}]{2021NatRP...3..344B}
{Bailes} M.,  et~al., 2021, \mn@doi [Nature Reviews Physics]
  {10.1038/s42254-021-00303-8}, \href
  {https://ui.adsabs.harvard.edu/abs/2021NatRP...3..344B} {3, 344}

\bibitem[\protect\citeauthoryear{{Barack} et~al.,}{{Barack}
  et~al.}{2019}]{2019CQGra..36n3001B}
{Barack} L.,  et~al., 2019, \mn@doi [Classical and Quantum Gravity]
  {10.1088/1361-6382/ab0587}, \href
  {https://ui.adsabs.harvard.edu/abs/2019CQGra..36n3001B} {36, 143001}

\bibitem[\protect\citeauthoryear{Biesiada}{Biesiada}{2018}]{Biesiada:2018nki}
Biesiada M.,  2018, in {38th Polish Astronomical Society Assembly}. pp 21--26

\bibitem[\protect\citeauthoryear{{B{\l}aut}, {Babak}  \&
  {Kr{\'o}lak}}{{B{\l}aut} et~al.}{2010}]{2010PhRvD..81f3008B}
{B{\l}aut} A.,  {Babak} S.,   {Kr{\'o}lak} A.,  2010, \mn@doi [\prd]
  {10.1103/PhysRevD.81.063008}, \href
  {https://ui.adsabs.harvard.edu/abs/2010PhRvD..81f3008B} {81, 063008}

\bibitem[\protect\citeauthoryear{{Bottke}, {Durda}, {Nesvorn{\'y}}, {Jedicke},
  {Morbidelli}, {Vokrouhlick{\'y}}  \& {Levison}}{{Bottke}
  et~al.}{2005}]{2005Icar..179...63B}
{Bottke} W.~F.,  {Durda} D.~D.,  {Nesvorn{\'y}} D.,  {Jedicke} R.,
  {Morbidelli} A.,  {Vokrouhlick{\'y}} D.,   {Levison} H.~F.,  2005, \mn@doi
  [\icarus] {10.1016/j.icarus.2005.05.017}, \href
  {https://ui.adsabs.harvard.edu/abs/2005Icar..179...63B} {179, 63}

\bibitem[\protect\citeauthoryear{Braginsky, Ryazhskaya  \&
  Vyatchanin}{Braginsky et~al.}{2006}]{BRAGINSKY20061}
Braginsky V.,  Ryazhskaya O.,   Vyatchanin S.,  2006, \mn@doi [Physics Letters
  A] {https://doi.org/10.1016/j.physleta.2005.09.073}, 350, 1

\bibitem[\protect\citeauthoryear{Britt, Yeomans, Housen  \& Consolmagno}{Britt
  et~al.}{2003}]{britt2003asteroid}
Britt D.~T.,  Yeomans D.,  Housen K.,   Consolmagno G.,  2003

\bibitem[\protect\citeauthoryear{Brownlee}{Brownlee}{2014}]{brownlee2014stardust}
Brownlee D.,  2014, Annual Review of Earth and Planetary Sciences, 42, 179

\bibitem[\protect\citeauthoryear{Campins, de Le{\'o}n, Licandro, Hendrix,
  S{\'a}nchez  \& Ali-Lagoa}{Campins et~al.}{2018}]{campins2018compositional}
Campins H.,  de Le{\'o}n J.,  Licandro J.,  Hendrix A.,  S{\'a}nchez J.~A.,
  Ali-Lagoa V.,  2018, in , Primitive Meteorites and Asteroids.
Elsevier, pp 345--369

\bibitem[\protect\citeauthoryear{Carry}{Carry}{2012}]{carry2012density}
Carry B.,  2012, Planetary and Space Science, 73, 98

\bibitem[\protect\citeauthoryear{{Chaibi}, {Geiger}, {Canuel}, {Bertoldi},
  {Landragin}  \& {Bouyer}}{{Chaibi} et~al.}{2016}]{2016PhRvD..93b1101C}
{Chaibi} W.,  {Geiger} R.,  {Canuel} B.,  {Bertoldi} A.,  {Landragin} A.,
  {Bouyer} P.,  2016, \mn@doi [\prd] {10.1103/PhysRevD.93.021101}, \href
  {https://ui.adsabs.harvard.edu/abs/2016PhRvD..93b1101C} {93, 021101}

\bibitem[\protect\citeauthoryear{{Chauvineau}, {Pireaux}  \&
  {Regimbau}}{{Chauvineau} et~al.}{2007}]{2007CQGra..24.3005C}
{Chauvineau} B.,  {Pireaux} S.,   {Regimbau} T.,  2007, \mn@doi [Classical and
  Quantum Gravity] {10.1088/0264-9381/24/11/013}, \href
  {https://ui.adsabs.harvard.edu/abs/2007CQGra..24.3005C} {24, 3005}

\bibitem[\protect\citeauthoryear{{Cheng} et~al.,}{{Cheng}
  et~al.}{2018}]{2018P&SS..157..104C}
{Cheng} A.~F.,  et~al., 2018, \mn@doi [\planss] {10.1016/j.pss.2018.02.015},
  \href {https://ui.adsabs.harvard.edu/abs/2018P&SS..157..104C} {157, 104}

\bibitem[\protect\citeauthoryear{Clark}{Clark}{1995}]{clark1995spectral}
Clark B.~E.,  1995, Journal of Geophysical Research: Planets, 100, 14443

\bibitem[\protect\citeauthoryear{Clement, Raymond  \& Kaib}{Clement
  et~al.}{2019}]{Clement_2019}
Clement M.~S.,  Raymond S.~N.,   Kaib N.~A.,  2019, \mn@doi [The Astronomical
  Journal] {10.3847/1538-3881/aaf21e}, 157, 38

\bibitem[\protect\citeauthoryear{{Cuzzi}, {Hogan}  \& {Bottke}}{{Cuzzi}
  et~al.}{2010}]{2010Icar..208..518C}
{Cuzzi} J.~N.,  {Hogan} R.~C.,   {Bottke} W.~F.,  2010, \mn@doi [\icarus]
  {10.1016/j.icarus.2010.03.005}, \href
  {https://ui.adsabs.harvard.edu/abs/2010Icar..208..518C} {208, 518}

\bibitem[\protect\citeauthoryear{{DeMeo} \& {Carry}}{{DeMeo} \&
  {Carry}}{2013}]{2013Icar..226..723D}
{DeMeo} F.~E.,  {Carry} B.,  2013, \mn@doi [\icarus]
  {10.1016/j.icarus.2013.06.027}, \href
  {https://ui.adsabs.harvard.edu/abs/2013Icar..226..723D} {226, 723}

\bibitem[\protect\citeauthoryear{DeMeo \& Carry}{DeMeo \&
  Carry}{2014}]{demeo2014solar}
DeMeo F.~E.,  Carry B.,  2014, Nature, 505, 629

\bibitem[\protect\citeauthoryear{{DeMeo}, {Binzel}, {Slivan}  \& {Bus}}{{DeMeo}
  et~al.}{2009}]{2009Icar..202..160D}
{DeMeo} F.~E.,  {Binzel} R.~P.,  {Slivan} S.~M.,   {Bus} S.~J.,  2009, \mn@doi
  [\icarus] {10.1016/j.icarus.2009.02.005}, \href
  {https://ui.adsabs.harvard.edu/abs/2009Icar..202..160D} {202, 160}

\bibitem[\protect\citeauthoryear{{DeMeo}, {Alexander}, {Walsh}, {Chapman}  \&
  {Binzel}}{{DeMeo} et~al.}{2015}]{2015aste.book...13D}
{DeMeo} F.~E.,  {Alexander} C.~M.~O.,  {Walsh} K.~J.,  {Chapman} C.~R.,
  {Binzel} R.~P.,  2015, {The Compositional Structure of the Asteroid Belt}.
pp 13--41, \mn@doi{10.2458/azu_uapress_9780816532131-ch002}

\bibitem[\protect\citeauthoryear{{Dell'Oro}, {Cellino}  \&
  {Paolicchi}}{{Dell'Oro} et~al.}{2012}]{2012MNRAS.425.1492D}
{Dell'Oro} A.,  {Cellino} A.,   {Paolicchi} P.,  2012, \mn@doi [\mnras]
  {10.1111/j.1365-2966.2012.21643.x}, \href
  {https://ui.adsabs.harvard.edu/abs/2012MNRAS.425.1492D} {425, 1492}

\bibitem[\protect\citeauthoryear{{Dohnanyi}}{{Dohnanyi}}{1969}]{1969JGR....74.2531D}
{Dohnanyi} J.~S.,  1969, \mn@doi [\jgr] {10.1029/JB074i010p02531}, \href
  {https://ui.adsabs.harvard.edu/abs/1969JGR....74.2531D} {74, 2531}

\bibitem[\protect\citeauthoryear{Farinella \& Davis}{Farinella \&
  Davis}{1992}]{farinella1992collision}
Farinella P.,  Davis D.~R.,  1992, \icarus, 97, 111

\bibitem[\protect\citeauthoryear{{Fedderke}, {Graham}  \&
  {Rajendran}}{{Fedderke} et~al.}{2021}]{2021PhRvD.103j3017F}
{Fedderke} M.~A.,  {Graham} P.~W.,   {Rajendran} S.,  2021, \mn@doi [\prd]
  {10.1103/PhysRevD.103.103017}, \href
  {https://ui.adsabs.harvard.edu/abs/2021PhRvD.103j3017F} {103, 103017}

\bibitem[\protect\citeauthoryear{Flanagan \& Hughes}{Flanagan \&
  Hughes}{2005}]{Flanagan_2005}
Flanagan {\'{E}}.~{\'{E}}.,  Hughes S.~A.,  2005, \mn@doi [New Journal of
  Physics] {10.1088/1367-2630/7/1/204}, 7, 204

\bibitem[\protect\citeauthoryear{{Gladman} et~al.,}{{Gladman}
  et~al.}{2009}]{2009Icar..202..104G}
{Gladman} B.~J.,  et~al., 2009, \mn@doi [\icarus]
  {10.1016/j.icarus.2009.02.012}, \href
  {https://ui.adsabs.harvard.edu/abs/2009Icar..202..104G} {202, 104}

\bibitem[\protect\citeauthoryear{Glassmeier, Boehnhardt, Koschny, K{\"u}hrt  \&
  Richter}{Glassmeier et~al.}{2007}]{glassmeier2007rosetta}
Glassmeier K.-H.,  Boehnhardt H.,  Koschny D.,  K{\"u}hrt E.,   Richter I.,
  2007, Space Science Reviews, 128, 1

\bibitem[\protect\citeauthoryear{Harry, Fritschel, Shaddock, Folkner  \&
  Phinney}{Harry et~al.}{2006}]{harry2006laser}
Harry G.~M.,  Fritschel P.,  Shaddock D.~A.,  Folkner W.,   Phinney E.~S.,
  2006, Classical and Quantum Gravity, 23, 4887

\bibitem[\protect\citeauthoryear{{Izidoro}, {Raymond}, {Morbidelli}  \&
  {Winter}}{{Izidoro} et~al.}{2015}]{2015MNRAS.453.3619I}
{Izidoro} A.,  {Raymond} S.~N.,  {Morbidelli} A.~r.,   {Winter} O.~C.,  2015,
  \mn@doi [\mnras] {10.1093/mnras/stv1835}, \href
  {https://ui.adsabs.harvard.edu/abs/2015MNRAS.453.3619I} {453, 3619}

\bibitem[\protect\citeauthoryear{Izidoro, Raymond, Pierens, Morbidelli, Winter
  \& Nesvorny}{Izidoro et~al.}{2016}]{izidoro2016asteroid}
Izidoro A.,  Raymond S.~N.,  Pierens A.,  Morbidelli A.,  Winter O.~C.,
  Nesvorny D.,  2016, The Astrophysical Journal, 833, 40

\bibitem[\protect\citeauthoryear{{Jedicke}, {Larsen}  \& {Spahr}}{{Jedicke}
  et~al.}{2002}]{2002aste.book...71J}
{Jedicke} R.,  {Larsen} J.,   {Spahr} T.,  2002, {Observational Selection
  Effects in Asteroid Surveys}.
pp 71--87

\bibitem[\protect\citeauthoryear{Jewitt, Weaver, Agarwal, Mutchler  \&
  Drahus}{Jewitt et~al.}{2010}]{jewitt2010recent}
Jewitt D.,  Weaver H.,  Agarwal J.,  Mutchler M.,   Drahus M.,  2010, Nature,
  467, 817

\bibitem[\protect\citeauthoryear{{Kaloshin}, {Zhang}  \& {Zhang}}{{Kaloshin}
  et~al.}{2015}]{2015arXiv151104835K}
{Kaloshin} V.,  {Zhang} J.,   {Zhang} K.,  2015, arXiv e-prints, \href
  {https://ui.adsabs.harvard.edu/abs/2015arXiv151104835K} {p. arXiv:1511.04835}

\bibitem[\protect\citeauthoryear{Kawamura et~al.,}{Kawamura
  et~al.}{2006}]{kawamura2006japanese}
Kawamura S.,  et~al., 2006, Classical and Quantum Gravity, 23, S125

\bibitem[\protect\citeauthoryear{Kawamura et~al.,}{Kawamura
  et~al.}{2011}]{kawamura2011japanese}
Kawamura S.,  et~al., 2011, Classical and Quantum Gravity, 28, 094011

\bibitem[\protect\citeauthoryear{{Kessler}}{{Kessler}}{1971}]{1971NASSP.267..595K}
{Kessler} D.~J.,  1971, {Estimate of Particle Densities and Collision Danger
  for Spacecraft Moving Through the Asteroid Belt}.
p.~595

\bibitem[\protect\citeauthoryear{{Kocsis}, {G{\'a}sp{\'a}r}  \&
  {M{\'a}rka}}{{Kocsis} et~al.}{2006}]{2006ApJ...648..411K}
{Kocsis} B.,  {G{\'a}sp{\'a}r} M.~E.,   {M{\'a}rka} S.,  2006, \mn@doi [\apj]
  {10.1086/505641}, \href
  {https://ui.adsabs.harvard.edu/abs/2006ApJ...648..411K} {648, 411}

\bibitem[\protect\citeauthoryear{{Korol} et~al.,}{{Korol}
  et~al.}{2020}]{2020A&A...638A.153K}
{Korol} V.,  et~al., 2020, \mn@doi [\aap] {10.1051/0004-6361/202037764}, \href
  {https://ui.adsabs.harvard.edu/abs/2020A&A...638A.153K} {638, A153}

\bibitem[\protect\citeauthoryear{{Krasinsky}, {Pitjeva}, {Vasilyev}  \&
  {Yagudina}}{{Krasinsky} et~al.}{2002}]{2002Icar..158...98K}
{Krasinsky} G.~A.,  {Pitjeva} E.~V.,  {Vasilyev} M.~V.,   {Yagudina} E.~I.,
  2002, \mn@doi [\icarus] {10.1006/icar.2002.6837}, \href
  {https://ui.adsabs.harvard.edu/abs/2002Icar..158...98K} {158, 98}

\bibitem[\protect\citeauthoryear{{LISA Study Team} et~al.}{{LISA Study Team}
  et~al.}{2000}]{lisa2000lisa}
{LISA Study Team} et~al., 2000, ESA System and Technology Study Report ESA-SCI,
  11

\bibitem[\protect\citeauthoryear{{Luo} et~al.}{{Luo}
  et~al.}{2016}]{2016CQGra..33c5010L}
{Luo} J.,  et~al., 2016, \mn@doi [Classical and Quantum Gravity]
  {10.1088/0264-9381/33/3/035010}, \href
  {https://ui.adsabs.harvard.edu/abs/2016CQGra..33c5010L} {33, 035010}

\bibitem[\protect\citeauthoryear{{Luo}, {Wang}, {Wu}, {Hu}  \& {Jin}}{{Luo}
  et~al.}{2021}]{2021PTEP.2021eA108L}
{Luo} Z.,  {Wang} Y.,  {Wu} Y.,  {Hu} W.,   {Jin} G.,  2021, \mn@doi [Progress
  of Theoretical and Experimental Physics] {10.1093/ptep/ptaa083}, \href
  {https://ui.adsabs.harvard.edu/abs/2021PTEP.2021eA108L} {2021, 05A108}

\bibitem[\protect\citeauthoryear{{Mainzer} et~al.,}{{Mainzer}
  et~al.}{2011}]{2011ApJ...731...53M}
{Mainzer} A.,  et~al., 2011, \mn@doi [\apj] {10.1088/0004-637X/731/1/53}, \href
  {https://ui.adsabs.harvard.edu/abs/2011ApJ...731...53M} {731, 53}

\bibitem[\protect\citeauthoryear{{Mapelli}}{{Mapelli}}{2021}]{2021hgwa.bookE...4M}
{Mapelli} M.,  2021, {Formation Channels of Single and Binary Stellar-Mass
  Black Holes}.
p.~4, \mn@doi{10.1007/978-981-15-4702-7\_16-1}

\bibitem[\protect\citeauthoryear{Marchis, Descamps, Baek, Harris, Kaasalainen,
  Berthier, Hestroffer  \& Vachier}{Marchis et~al.}{2008}]{MARCHIS200897}
Marchis F.,  Descamps P.,  Baek M.,  Harris A.,  Kaasalainen M.,  Berthier J.,
  Hestroffer D.,   Vachier F.,  2008, \mn@doi [Icarus]
  {https://doi.org/10.1016/j.icarus.2008.03.007}, 196, 97

\bibitem[\protect\citeauthoryear{Margot, Nolan, Benner, Ostro, Jurgens,
  Giorgini, Slade  \& Campbell}{Margot et~al.}{2002}]{margot2002binary}
Margot J.-L.,  Nolan M.,  Benner L.,  Ostro S.,  Jurgens R.,  Giorgini J.,
  Slade M.,   Campbell D.,  2002, Science, 296, 1445

\bibitem[\protect\citeauthoryear{{Margot}, {Pravec}, {Taylor}, {Carry}  \&
  {Jacobson}}{{Margot} et~al.}{2015}]{2015aste.book..355M}
{Margot} J.~L.,  {Pravec} P.,  {Taylor} P.,  {Carry} B.,   {Jacobson} S.,
  2015, {Asteroid Systems: Binaries, Triples, and Pairs}.
pp 355--374, \mn@doi{10.2458/azu_uapress_9780816532131-ch019}

\bibitem[\protect\citeauthoryear{{Matlack} \& {Metz}}{{Matlack} \&
  {Metz}}{1967}]{LOSALAMOS}
{Matlack} G.,  {Metz} C.,  1967, Radiation Characteristics of Plutonium-238,
  \url {https://fas.org/sgp/othergov/doe/lanl/docs1/00314834.pdf}

\bibitem[\protect\citeauthoryear{{Mei} et~al.}{{Mei}
  et~al.}{2021}]{2021PTEP.2021eA107M}
{Mei} J.,  et~al., 2021, \mn@doi [Progress of Theoretical and Experimental
  Physics] {10.1093/ptep/ptaa114}, \href
  {https://ui.adsabs.harvard.edu/abs/2021PTEP.2021eA107M} {2021, 05A107}

\bibitem[\protect\citeauthoryear{Men, Ni  \& Wang}{Men
  et~al.}{2010}]{men2010design}
Men J.-r.,  Ni W.-t.,   Wang G.,  2010, Chinese Astronomy and Astrophysics, 34,
  434

\bibitem[\protect\citeauthoryear{Miao, Yang  \& Martynov}{Miao
  et~al.}{2018}]{miao2018towards}
Miao H.,  Yang H.,   Martynov D.,  2018, Physical Review D, 98, 044044

\bibitem[\protect\citeauthoryear{{Michel}, {DeMeo}  \& {Bottke}}{{Michel}
  et~al.}{2015}]{2015aste.book.....M}
{Michel} P.,  {DeMeo} F.~E.,   {Bottke} W.~F.,  2015, {Asteroids IV},
  \mn@doi{10.2458/azu_uapress_9780816532131.
}

\bibitem[\protect\citeauthoryear{{Moons}}{{Moons}}{1996}]{1996CeMDA..65..175M}
{Moons} M.,  1996, \mn@doi [Celestial Mechanics and Dynamical Astronomy]
  {10.1007/BF00048446}, \href
  {https://ui.adsabs.harvard.edu/abs/1996CeMDA..65..175M} {65, 175}

\bibitem[\protect\citeauthoryear{{NASA/McREL}}{{NASA/McREL}}{2007}]{DawnPhotoJ}
{NASA/McREL} 2007, PIA19380: Asteroid Belt, \url
  {https://photojournal.jpl.nasa.gov/catalog/PIA19380}

\bibitem[\protect\citeauthoryear{{Naidu} et~al.,}{{Naidu}
  et~al.}{2020}]{2020Icar..34813777N}
{Naidu} S.~P.,  et~al., 2020, \mn@doi [\icarus] {10.1016/j.icarus.2020.113777},
  \href {https://ui.adsabs.harvard.edu/abs/2020Icar..34813777N} {348, 113777}

\bibitem[\protect\citeauthoryear{{Nesvorn{\'y}}, {Bro{\v{z}}}  \&
  {Carruba}}{{Nesvorn{\'y}} et~al.}{2015}]{2015aste.book..297N}
{Nesvorn{\'y}} D.,  {Bro{\v{z}}} M.,   {Carruba} V.,  2015, {Identification and
  Dynamical Properties of Asteroid Families}.
pp 297--321, \mn@doi{10.2458/azu_uapress_9780816532131-ch016}

\bibitem[\protect\citeauthoryear{{Ni}}{{Ni}}{2009}]{2009CQGra..26g5021N}
{Ni} W.-T.,  2009, \mn@doi [Classical and Quantum Gravity]
  {10.1088/0264-9381/26/7/075021}, \href
  {https://ui.adsabs.harvard.edu/abs/2009CQGra..26g5021N} {26, 075021}

\bibitem[\protect\citeauthoryear{Ni}{Ni}{2013}]{ni2013astrod}
Ni W.-T.,  2013, International Journal of Modern Physics D, 22, 1341004

\bibitem[\protect\citeauthoryear{{Ni}}{{Ni}}{2016}]{2016IJMPD..2530001N}
{Ni} W.-T.,  2016, \mn@doi [International Journal of Modern Physics D]
  {10.1142/S0218271816300019}, \href
  {https://ui.adsabs.harvard.edu/abs/2016IJMPD..2530001N} {25, 1630001}

\bibitem[\protect\citeauthoryear{{Ni} et~al.,}{{Ni}
  et~al.}{2010}]{2010cosp...38.3821N}
{Ni} W.~T.,  et~al., 2010, in 38th COSPAR Scientific Assembly. p.~11

\bibitem[\protect\citeauthoryear{Petit, Chambers, Franklin  \& Nagasawa}{Petit
  et~al.}{2002}]{petit2002primordial}
Petit J.-M.,  Chambers J.,  Franklin F.,   Nagasawa M.,  2002, Asteroids III,
  711

\bibitem[\protect\citeauthoryear{{Pravec} \& {Harris}}{{Pravec} \&
  {Harris}}{2007}]{2007Icar..190..250P}
{Pravec} P.,  {Harris} A.~W.,  2007, \mn@doi [\icarus]
  {10.1016/j.icarus.2007.02.023}, \href
  {https://ui.adsabs.harvard.edu/abs/2007Icar..190..250P} {190, 250}

\bibitem[\protect\citeauthoryear{Pravec et~al.,}{Pravec
  et~al.}{2016}]{pravec2016binary}
Pravec P.,  et~al., 2016, Icarus, 267, 267

\bibitem[\protect\citeauthoryear{{Reitze} et~al.,}{{Reitze}
  et~al.}{2019}]{2019BAAS...51g..35R}
{Reitze} D.,  et~al., 2019, in \baas. p.~35 (\mn@eprint {arXiv} {1907.04833})

\bibitem[\protect\citeauthoryear{{Robson}, {Cornish}  \& {Liu}}{{Robson}
  et~al.}{2019}]{2019CQGra..36j5011R}
{Robson} T.,  {Cornish} N.~J.,   {Liu} C.,  2019, \mn@doi [Classical and
  Quantum Gravity] {10.1088/1361-6382/ab1101}, \href
  {https://ui.adsabs.harvard.edu/abs/2019CQGra..36j5011R} {36, 105011}

\bibitem[\protect\citeauthoryear{{Ruan}, {Guo}, {Cai}  \& {Zhang}}{{Ruan}
  et~al.}{2020}]{2020IJMPA..3550075R}
{Ruan} W.-H.,  {Guo} Z.-K.,  {Cai} R.-G.,   {Zhang} Y.-Z.,  2020, \mn@doi
  [International Journal of Modern Physics A] {10.1142/S0217751X2050075X},
  \href {https://ui.adsabs.harvard.edu/abs/2020IJMPA..3550075R} {35, 2050075}

\bibitem[\protect\citeauthoryear{{Ruiter}, {Belczynski}, {Benacquista},
  {Larson}  \& {Williams}}{{Ruiter} et~al.}{2010}]{2010ApJ...717.1006R}
{Ruiter} A.~J.,  {Belczynski} K.,  {Benacquista} M.,  {Larson} S.~L.,
  {Williams} G.,  2010, \mn@doi [\apj] {10.1088/0004-637X/717/2/1006}, \href
  {https://ui.adsabs.harvard.edu/abs/2010ApJ...717.1006R} {717, 1006}

\bibitem[\protect\citeauthoryear{Russell}{Russell}{2012}]{russell2012galileo}
Russell C.~T.,  2012, The Galileo Mission.
Springer Science \& Business Media

\bibitem[\protect\citeauthoryear{{Russell} \& {Raymond}}{{Russell} \&
  {Raymond}}{2011}]{2011SSRv..163....3R}
{Russell} C.~T.,  {Raymond} C.~A.,  2011, \mn@doi [\ssr]
  {10.1007/s11214-011-9836-2}, \href
  {https://ui.adsabs.harvard.edu/abs/2011SSRv..163....3R} {163, 3}

\bibitem[\protect\citeauthoryear{{Ryan}, {Mizuno}, {Shenoy}, {Woodward},
  {Carey}, {Noriega-Crespo}, {Kraemer}  \& {Price}}{{Ryan}
  et~al.}{2015}]{2015A&A...578A..42R}
{Ryan} E.~L.,  {Mizuno} D.~R.,  {Shenoy} S.~S.,  {Woodward} C.~E.,  {Carey}
  S.~J.,  {Noriega-Crespo} A.,  {Kraemer} K.~E.,   {Price} S.~D.,  2015,
  \mn@doi [\aap] {10.1051/0004-6361/201321375}, \href
  {https://ui.adsabs.harvard.edu/abs/2015A&A...578A..42R} {578, A42}

\bibitem[\protect\citeauthoryear{{Sathyaprakash} \& {Schutz}}{{Sathyaprakash}
  \& {Schutz}}{2009}]{2009LRR....12....2S}
{Sathyaprakash} B.~S.,  {Schutz} B.~F.,  2009, \mn@doi [Living Reviews in
  Relativity] {10.12942/lrr-2009-2}, \href
  {https://ui.adsabs.harvard.edu/abs/2009LRR....12....2S} {12, 2}

\bibitem[\protect\citeauthoryear{Sato et~al.,}{Sato
  et~al.}{2017}]{sato2017status}
Sato S.,  et~al., 2017, in Journal of Physics: Conference Series. p. 012010

\bibitem[\protect\citeauthoryear{{Schneider}, {Ferrari}, {Matarrese}  \&
  {Portegies Zwart}}{{Schneider} et~al.}{2001}]{2001MNRAS.324..797S}
{Schneider} R.,  {Ferrari} V.,  {Matarrese} S.,   {Portegies Zwart} S.~F.,
  2001, \mn@doi [\mnras] {10.1046/j.1365-8711.2001.04217.x}, \href
  {https://ui.adsabs.harvard.edu/abs/2001MNRAS.324..797S} {324, 797}

\bibitem[\protect\citeauthoryear{Schumaker}{Schumaker}{2003}]{schumaker2003disturbance}
Schumaker B.~L.,  2003, Classical and Quantum Gravity, 20, S239

\bibitem[\protect\citeauthoryear{Schutz}{Schutz}{1999}]{schutz1999gravitational}
Schutz B.~F.,  1999, Classical and Quantum Gravity, 16, A131

\bibitem[\protect\citeauthoryear{{Sesana} et~al.,}{{Sesana}
  et~al.}{2019}]{2019arXiv190811391S}
{Sesana} A.,  et~al., 2019, arXiv e-prints, \href
  {https://ui.adsabs.harvard.edu/abs/2019arXiv190811391S} {p. arXiv:1908.11391}

\bibitem[\protect\citeauthoryear{{Shaddock}}{{Shaddock}}{2004}]{2004PhRvD..69b2001S}
{Shaddock} D.~A.,  2004, \mn@doi [\prd] {10.1103/PhysRevD.69.022001}, \href
  {https://ui.adsabs.harvard.edu/abs/2004PhRvD..69b2001S} {69, 022001}

\bibitem[\protect\citeauthoryear{Snodgrass et~al.,}{Snodgrass
  et~al.}{2010}]{snodgrass2010collision}
Snodgrass C.,  et~al., 2010, Nature, 467, 814

\bibitem[\protect\citeauthoryear{Stozhkov, Svirzhevsky  \& Makhmutov}{Stozhkov
  et~al.}{2001}]{stozhkov2001cosmic}
Stozhkov Y.~I.,  Svirzhevsky N.,   Makhmutov V.,  2001

\bibitem[\protect\citeauthoryear{Sullivan, Veske, M{\'{a}}rka, Bartos, Ballmer,
  Shawhan  \& M{\'{a}}rka}{Sullivan et~al.}{2020}]{Sullivan_2020}
Sullivan A.~G.,  Veske D.,  M{\'{a}}rka Z.,  Bartos I.,  Ballmer S.,  Shawhan
  P.,   M{\'{a}}rka S.,  2020, \mn@doi [Classical and Quantum Gravity]
  {10.1088/1361-6382/abb260}, 37, 205005

\bibitem[\protect\citeauthoryear{{Tedesco} \& {Desert}}{{Tedesco} \&
  {Desert}}{2002}]{2002AJ....123.2070T}
{Tedesco} E.~F.,  {Desert} F.-X.,  2002, \mn@doi [\aj] {10.1086/339482}, \href
  {https://ui.adsabs.harvard.edu/abs/2002AJ....123.2070T} {123, 2070}

\bibitem[\protect\citeauthoryear{{Tedesco}, {Cellino}  \&
  {Zappal{\'a}}}{{Tedesco} et~al.}{2005}]{2005AJ....129.2869T}
{Tedesco} E.~F.,  {Cellino} A.,   {Zappal{\'a}} V.,  2005, \mn@doi [\aj]
  {10.1086/429734}, \href
  {https://ui.adsabs.harvard.edu/abs/2005AJ....129.2869T} {129, 2869}

\bibitem[\protect\citeauthoryear{{Tinto} \& {Dhurandhar}}{{Tinto} \&
  {Dhurandhar}}{2014}]{2014LRR....17....6T}
{Tinto} M.,  {Dhurandhar} S.~V.,  2014, \mn@doi [Living Reviews in Relativity]
  {10.12942/lrr-2014-6}, \href
  {https://ui.adsabs.harvard.edu/abs/2014LRR....17....6T} {17, 6}

\bibitem[\protect\citeauthoryear{{Tinto} \& {Dhurandhar}}{{Tinto} \&
  {Dhurandhar}}{2021}]{2021LRR....24....1T}
{Tinto} M.,  {Dhurandhar} S.~V.,  2021, \mn@doi [Living Reviews in Relativity]
  {10.1007/s41114-020-00029-6}, \href
  {https://ui.adsabs.harvard.edu/abs/2021LRR....24....1T} {24, 1}

\bibitem[\protect\citeauthoryear{{Tricarico}}{{Tricarico}}{2009}]{2009CQGra..26h5003T}
{Tricarico} P.,  2009, \mn@doi [Classical and Quantum Gravity]
  {10.1088/0264-9381/26/8/085003}, \href
  {https://ui.adsabs.harvard.edu/abs/2009CQGra..26h5003T} {26, 085003}

\bibitem[\protect\citeauthoryear{Veverka et~al.,}{Veverka
  et~al.}{2001}]{veverka2001landing}
Veverka J.,  et~al., 2001, Nature, 413, 390

\bibitem[\protect\citeauthoryear{{Vinet}}{{Vinet}}{2006}]{2006CQGra..23.4939V}
{Vinet} J.-Y.,  2006, \mn@doi [Classical and Quantum Gravity]
  {10.1088/0264-9381/23/15/012}, \href
  {https://ui.adsabs.harvard.edu/abs/2006CQGra..23.4939V} {23, 4939}

\bibitem[\protect\citeauthoryear{{Walsh} \& {Jacobson}}{{Walsh} \&
  {Jacobson}}{2015}]{2015aste.book..375W}
{Walsh} K.~J.,  {Jacobson} S.~A.,  2015, {Formation and Evolution of Binary
  Asteroids}.
pp 375--393, \mn@doi{10.2458/azu_uapress_9780816532131-ch020}

\bibitem[\protect\citeauthoryear{{Wetherill}}{{Wetherill}}{1967}]{1967JGR....72.2429W}
{Wetherill} G.~W.,  1967, \mn@doi [\jgr] {10.1029/JZ072i009p02429}, \href
  {https://ui.adsabs.harvard.edu/abs/1967JGR....72.2429W} {72, 2429}

\bibitem[\protect\citeauthoryear{Wilkening}{Wilkening}{1977}]{wilkening19778}
Wilkening L.,  1977, in International Astronomical Union Colloquium. pp
  389--396

\makeatother
\end{thebibliography}

\end{document}